\documentclass[twocolumn]{aa}  
\usepackage{epsfig}
\def\simlt{\ \raise -2.truept\hbox{\rlap{\hbox{$\sim$}}\raise5.truept   %
\hbox{$<$}\ }}
\def\simgt{\ \raise -2.truept\hbox{\rlap{\hbox{$\sim$}}\raise5.truept   %
\hbox{$>$}\ }}                                                          %

\def\be{\begin{equation}}
\def\ee{\end{equation}}
\def\newline{\hfil\break}

\def\la{\mathrel{\hbox{\rlap{\hbox{\lower4pt\hbox{$\sim$}}}\hbox{$<$}}}}
\def\ga{\mathrel{\hbox{\rlap{\hbox{\lower4pt\hbox{$\sim$}}}\hbox{$>$}}}}

\begin{document}
\title{On the ability of spectroscopic SZ effect measurements to determine
the temperature structure of galaxy clusters}
   \author{S. Colafrancesco\inst{1,2} and P. Marchegiani\inst{2,3}}
   \offprints{S. Colafrancesco}
   \institute{   ASI-ASDC
              c/o ESRIN, Via G. Galilei snc, I-00040 Frascati, Italy
              Email: Sergio.Colafrancesco@asi.it
   \and
              INAF - Osservatorio Astronomico di Roma
              via Frascati 33, I-00040 Monteporzio, Italy.
              Email: cola@mporzio.astro.it
   \and
              Dipartimento di Fisica, Universit\`a di Roma La Sapienza, P.le A. Moro 2, Roma, Italy
              Email: marchegiani@mporzio.astro.it
             }
\date{Received  / Accepted  }
\authorrunning {S. Colafrancesco and P. Marchegiani}
\titlerunning {SZ effect and cluster temperature profiles}
%
\abstract
  {}
   {
   We explore in this paper the ability of spatially resolved spectroscopic measurements
   of the SZ effect (SZE) to determine the temperature profile of galaxy clusters. We
   derive a general formalism for the thermal SZE in galaxy clusters
   with a non-uniform temperature profile that can be applied to
   both cool-core clusters and non-cool core cluster with an isothermal
   or non-isothermal temperature structure.
   }
  {
  We derive an inversion technique through which the electron distribution function
  can be extracted from spectroscopic SZE observations over a wide frequency
  range. We study the fitting procedure to extract the cluster temperature
  from a set of simulated spatially resolved spectroscopic SZE observations in
  different bands of the spectrum, from 100 to 450 GHz.
  }
   {
  The results of our analysis for three different cluster prototypes
  (A2199 with a low-temperature cool core, Perseus with a relatively
  high-temperature cool core, Ophiuchus with an isothermal temperature distribution)
  provide both the required precision of the SZE observations and the optimal frequency
  bands for a determination of the cluster temperature similar or better than
  that obtainable from X-ray observations. The precision of SZE-derived temperature
  is also discussed for the outer regions of clusters.
  We also study the possibility to extract, from our method, the
  parameters characterizing the non-thermal SZE spectrum of the relativistic plasma
  contained in the lobes of radio galaxies as well as the spectrum of relativistic
  electrons co-spatially distributed with the thermal plasma in clusters with non-thermal phenomena.}
  {We find that the next generation SZE experiments with
  spectroscopic capabilities, like those using FTS spectrometers
  with imaging capabilities, can provide precise temperature
  distribution measurements over a large range of radial distances
  for galaxy clusters even out to substantial redshifts.}


 \maketitle

\section{Introduction}
 \label{sec.intro}

Galaxy clusters are the largest gravitationally bound container of
diffuse baryons in the universe. These baryons accomodate in the
cluster gravitational potential well dominated by Cold Dark Matter
and show, in their hotter phase (i.e. the hot Intra Cluster Medium
-- hereafter ICM -- with temperature $T_e \sim 10^7 - 10^8$ K), a
complex temperature distribution in the cluster atmospheres as
indicated by observations (see, e.g., Arnaud 2005 for a review;
Pratt et al. 2007 for recent XMM-Newton results, Ehlert \& Ulmer 2009
for recent Chandra results, Sato et al. 2009 for recent Suzaku
results) and numerical simulations (see Borgani et al. 2008 for a
recent review).

Spatially resolved studies of the ICM temperature distribution in
galaxy clusters are crucial for both astrophysical and
cosmological applications since these studies provide information
on the thermodynamic state of the ICM, allow to measure the total
mass of these systems through the hydrostatic equilibrium equation
(see e.g. Arnaud 2005, Colafrancesco \& Giordano 2007), and allow
to set constraints on the relevant cosmological parameters
($\Omega_m, \Omega_{DE}, \Omega_b)$ through the study of the
cluster population evolution.

On the cosmological side, once the gas mass is determined from
X-ray observations in the deprojection or fitting analysis (see,
e.g., Fabian et al. 1981; Ettori \& Fabian 1999) and the total
mass is estimated through the hydrostatic equilibrium condition,
it is possible to derive the cluster gas mass fraction that can be
used to constrain the cosmological density parameter, if combined
with primordial nucleo-synthesis calculations (e.g., White et al.
1993, Ettori et al. 2009).
Cosmological applications of the cluster population evolution
require precise measurements of the cluster temperature and of the
correlation between temperature and other relevant physical
quantities, in order to describe correctly the cluster population
(total mass, luminosity, Compton parameter, number counts,
redshift distribution, luminosity function).
If well calibrated, the slope and evolution of cluster scaling
relations, such as gas mass versus temperature (e.g.,
Colafrancesco \& Vittorio 1994, Colafrancesco et al. 1994, 1997,
Voit 2000, Finoguenov, Reiprich, \& Boehringer 2001) and cluster
size versus temperature (e.g., Colafrancesco, et al. 1997, Mohr et
al. 2000; Verde et al. 2001), can also be used to constrain
cosmological and structure formation models.

On the astrophysical side, the precise determination of important
quantities such as entropy, pressure, and -- under the assumption
of hydrostatic equilibrium -- the total cluster mass, is dependent
on accurate estimation of the temperature profile.
Accurate temperature profiles are fundamental in determining the
gas entropy distribution (see Lloyd-Davies, Ponman, \& Cannon 2000
and references therein), which is a powerful tool to explore the
role of non-gravitational processes that could alter the specific
thermal energy in the ICM (Colafrancesco 2008a for a review, Kang
et al. 2007, Colafrancesco \& Giordano 2006, 2007), to assess the
structure of cool cores and their surrounding atmosphere in order
to probe the interplay of non-thermal and thermal particle
distributions in clusters (e.g., Colafrancesco, Dar \& DeRujula
2004, Guo \& Ho 2008, Colafrancesco \& Marchegiani 2008 and
references therein) and to assess the amount of non-thermal
(point-like and/or diffuse) emission mechanisms detectable in the
hard X-ray domain (see discussion in Colafrancesco \& Marchegiani
2009).

The standard methodology to recover the ICM temperature
distribution  makes use of high resolution spectroscopic X-ray
observations (XMM, Chandra, Suzaku and previously ROSAT, ASCA, and
BeppoSAX).
Such observations provide the key measurable characteristics of
the ICM, i.e., the temperature and density of this plasma. Because
of limited photon statistics it is usual to measure the density
and temperature in terms of radial profiles (see, e.g., Pratt et
al. 2007). However, while the density of
the ICM is relatively easy to measure from the surface brightness
profile of a galaxy cluster, precise temperature determination
requires high photon statistics to build, and fit, a spectrum.
Therefore, ICM temperature profiles are typically determined with
considerably less spatial resolution than density profiles. The
measurement of radial temperature profiles is further complicated
by the density squared dependence of the thermal bremsstrahlung
X-ray emissivity, $\varepsilon_{brem} \propto n^2_e(r)
T_e^{1/2}(r)$. The steep drop of the X-ray surface brightness with
distance from the centre, combined with the background from
cosmic, solar and instrumental sources, makes, hence, accurate
X-ray measurement of the temperature distribution at large
distances from the cluster centre a technically challenging task
(see discussion in Pratt et al. 2007).

ICM temperature can also be measured, alternatively, by using the
Inverse Compton scattering of CMB photons off thermal electrons
residing in the cluster atmosphere -- the Sunyaev-Zel'dovich
effect (SZE; see Sunyaev \& Zel'dovich 1980, Birkinshaw 1999,
Colafrancesco 2007 for reviews). This effect provides a CMB
temperature change
\begin{equation}
\frac{\Delta T}{T_0} = \frac{\sigma_T}{m_e c^2} \int_\ell d \ell n_e k_B T_e \cdot g(x)
\label{eq.szth}
\end{equation}
that depends on the cluster temperature $T_e$ directly from its
amplitude $\propto y = \frac{\sigma_T}{m_e c^2} \int_\ell d \ell
n_e k_B T_e$ (where $\sigma_T$ is the Thomson cross section, $m_e$
is the electron mass, $n_e$ is the electron density, $T_e$ is the
cluster temperature, the integral is performed along the line of
sight $\ell$), and from its spectrum $\propto g(x)$ (where $x \equiv
h \nu / k_B T_0$ is the frequency normalized to the CMB energy,
$h$ is the Planck constant, $k_B$ is the Boltzmann constant and
$T_0$ is the present-day CMB temperature).
The temperature dependence is explicitly included in the function
$g(x)$ through the relativistic effects of the Compton Scattering
that are more prominent in the high frequency region of the SZE
spectrum at $\nu \simgt 300$ GHz (see Birkinshaw 1999,
Colafrancesco, Marchegiani \& Palladino 2003, Colafrancesco 2007
for a review), where it is therefore possible to measure directly
the ICM temperature.
This is evident looking at the different temperature dependence of
the three basic spectral features that characterize the thermal
SZE signal:\\
i) the minimum in its intensity located at the frequency
 \begin{eqnarray}
 x_{th,min} & \approx &2.265(1-0.0927 \theta_e + 2.38 \theta_e^2) \nonumber \\
            & & + \tau (-0.00674 + 0.466 \theta_e) \; ,
 \label{eq.xmin}
 \end{eqnarray}
where $\theta_e \equiv k_B T_e / m_ec^2$, whose value depends
weakly on the electron spectrum (i.e. on $T_e$ and $n_e$) and
equals $\sim 2.26$;\\
ii) the crossover frequency, $x_0$, whose value depends on the
electron pressure/energy density and optical depth
 \be
 x_{th,0} \approx a(T_e) + \tau b(T_e)
 \label{eq.x0}
 \ee
with $a(T_e)= 3.830(1+1.162 \theta_e -0.8144 \theta_e^2)$ and
$b(T_e)= 3.021 \theta_e-8.672\theta_e^2$, and is found at a
frequency $> 3.83$ for increasing values of $T_e$ (the value
$x_{th,0}=3.83$ is found in the non-relativistic limit, or in the
limit $T_e \to 0$);\\
iii) the maximum of its intensity located at the frequency
 \begin{eqnarray}
 x_{th,max} & \approx & 6.511(1+2.41\theta_e -4.96\theta_e^2) \nonumber \\
            & & +\tau(0.0161+8.16\theta_e -35.9\theta_e^2)
 \label{eq.xmax}
 \end{eqnarray}
and depends sensitively on the nature of the electron population
and on its energy (momentum) spectrum (see Dolgov et al. 2001,
Colafrancesco et al. 2003, Colafrancesco et al. 2009 for the case
of electrons with a thermal spectrum; see also Colafrancesco et
al. 2003 and Colafrancesco 2004, Colafrancesco 2005, Colafrancesco
2007, Colafrancesco 2008b for the case of electron with different
spectra).

The SZE can be used, therefore, as an alternative probe of the ICM
temperature provided that detailed spectral measurements extending
out to high frequencies can be obtained (see Colafrancesco 2007
for a review, see Colafrancesco et al. 2009).\\
The linear dependence of the SZE temperature change $\Delta T$
from the electronic density $n_e$ and temperature $T_e$ (see eq.
\ref{eq.szth}) makes it possible -- in addition -- to enhance the
sensitivity of the SZE measurements in the outer regions of the
cluster while maintaining an adequate sensitivity also in the
inner parts.
Therefore SZE observations could provide a more uniform and
spatially extended coverage of the temperature profile in galaxy
clusters than X-ray observations.

Little or nothing is known about cluster temperature profiles
obtained directly from SZE observations. Present-day and planned
SZE observations from both ground based experiments (multi-band
bolometers and interferometers) or coming space-borne experiments
(e.g., OLIMPO and PLANCK) require complementary X-ray observations
to put tight constraints on the cluster temperature, even though a
somewhat more moderate precision on cluster temperature seems to
be sufficient to achieve a reasonable statistical knowledge on the
CDM power spectrum, namely on $\sigma_8$ (Juin et al. 2007).\\
We must also stress, in this context, that the use of X-ray
observed quantities in the SZE data analysis is, however, a
delicate procedure, because the SZE-derived temperature is
weighted by the Compton parameter $y$.
It is known, for instance, that
the derived peculiar velocity  is systematically shifted by $\sim
10-20 \%$ due to this reason (see e.g., Hansen 2004).\\
While observations of the SZE are becoming increasingly accurate
(e.g., LaRoque et al. 2002; De Petris et al. 2002; Battistelli et
al. 2003, Halverson et al. 2009, Staniszewski et al.
2009), most analyses of SZ data are still made under
the simplifying assumption of iso-thermality, or single
temperature plasma. Several studies have also considered the
ability of future SZ observations to extract the cluster physical
parameters (see, e.g., Knox, Holder \& Church 2004; Aghanim,
Hansen \& Lagache 2005), but always under the assumption of
iso-thermality.\\
A desirable solution with the SZE observations is to achieve
simultaneous good angular resolution and high sensitivity in order
to deproject the SZE signal properly throughout the whole cluster
atmosphere.

In this paper we discuss specifically the strategy that will be
more effective for obtaining spatially-resolved temperature
measurements which are only based on SZE observations. Such
strategy needs to overcome the following problems:
i) achieve spatially resolved spectral measurements of the cluster
temperature;
ii) measure the temperature profile over a wide spatial range in
order to disentangle the cool cores region from the outer cluster
atmosphere (or regions with different temperature in the overall
atmosphere);
iii) obtain detailed and independent temperature measurements
which require a wide frequency band coverage with sufficient
spectral resolution.
These points are the main issues that our study wants to address.

The outline of the paper is the following. We describe the general
theory of the SZE in Sect.2, and we generalize this theory to the
case of clusters with non-uniform T-profiles, like e.g. cool core
clusters, in Sect. 3. We describe a fitting procedure for
spatially resolved spectroscopic SZE observations in Sect.4, where
we first discuss the case of the thermal SZE. The analysis of the
non-thermal SZE and the combination of thermal and non-thermal SZE
are also discussed, for completeness, in Sect.5. The non-thermal
SZE is important both for the intrinsic study of non-thermal
and/or relativistic plasmas in clusters and for determining their
role as biases for the study of the thermal plasma (i.e. the ICM).
We present in Sect. 6 the application of our study for two
specific galaxy clusters: Perseus, which has a cool core, and
Ophiuchus, which is approximately isothermal. We discuss our
results and summarize our conclusions in the final Sect.7.

Throughout the paper, we use a flat, vacuum--dominated
cosmological model with $\Omega_m = 0.3$, $\Omega_{\Lambda} = 0.7$
and $H_0 = 70$ km s$^{-1}$ Mpc$^{-1}$.

\section{The SZE: theory}

In order to derive the expression of the SZE which is valid for a
galaxy cluster with spatially varying density and temperature,
we need to recall the general expressions for the SZE formulae.\\
We use here the formalism presented by Wright (1979), and
subsequently developed by Birkinshaw (1999), En\ss lin \& Kaiser
(2000), Colafrancesco, Marchegiani \& Palladino (2003).
We note that B\oe hm \& Lavalle (2009) recently proposed a
covariant formalism for the SZE, claiming, erroneously, that the
formalism presented in Wright (1979) and further developed by
other authors was incorrect. We have verified (both numerically
and analytically) that the two approaches, in the Thomson limit,
are fully equivalent and provide the same results (see Appendix A
for details). In addition to our verification, Nozawa \& Kohyama
(2009) have also shown analytically that the two approaches are
fully equivalent, pointing out the errors made by  B\oe hm \&
Lavalle (2009).

The analytic expression for the upscattered CMB spectrum is:
\begin{equation}
 \label{spettro_ris}
I(x)=\int_{-\infty}^{+\infty} I_0(xe^{-s}) P(s) ds
\end{equation}
(see e.g. Colafrancesco, Marchegiani \& Palladino 2003) where
\begin{equation} \label{spettro_inc}
I_0(x)=2\frac{(k_B T_0)^3}{(hc)^2} \frac{x^3}{e^x-1}
\end{equation}
with $x=h\nu/k_B T_0$.\\
The function $P(s)$ is given by the following expression
\begin{equation}
P(s)=\sum_{n=0}^{+\infty} \frac{e^{-\tau} \tau^n}{n!} P_n(s) \; ,
\end{equation}
where
\begin {equation} \label{tau}
\tau=\sigma_T \int_\ell n_e d\ell
\end{equation}
is the electron optical depth and
\begin{equation}
P_n(s)=\underbrace{P_1(s) \otimes \ldots \otimes P_1(s)}_{
 \mbox{n times}} \; \;,
 \label{eq.pns}
\end{equation}
where the symbol $\otimes$ indicates the convolution product, and
\begin{equation}
P_1(s)=\int_0^\infty f_e(p) P_s(s,p) dp \; .
 \label{eq.p1s}
\end{equation}
The function $f_e(p)$ is the electron momentum distribution
function (with $p=\beta \gamma$) and is normalized as to give
$\int_0^\infty f_e(p)dp=1$.
The function $P_s(s,p)$ is known from the basic physics of the
Compton scattering and is given by
\begin{equation}
P_s(s,p)ds=P(e^s,p)e^s ds \; ,
\end{equation}
with
\begin{eqnarray}
P(t,p)&=&-\frac{3|1-t|}{32p^6t}\left[1+(10+8p^2+4p^4)t+t^2\right]+
\nonumber \\
& & +\frac{3(1+t)}{8p^5}
\left[\frac{3+3p^2+p^4}{\sqrt{1+p^2}}+ \right.\nonumber \\
 & & \left.-\frac{3+2p^2}{2p} (2 \, \textrm{arcsinh} (p) -|\ln(t)|)\right] \; .
 \label{funz.ptp}
\end{eqnarray}
Our aim is to invert eq.(\ref{spettro_ris}) in order to derive the
spectral and spatial parameters of the electrons distribution from
measurements of the SZE only. We derive here an analytical
solution to this problem, while in Sect. 4 we present a numerical
procedure to solve it.

The approximated expression of $\Delta I(x) $ at first order in
$\tau$ is
\begin{equation}
\Delta I(x)=\tau [J_1(x) - I_0(x)]
\end{equation}
where
\begin{equation}
J_1(x)= \int_{-\infty}^{+\infty} I_0(xe^{-s}) P_1(s) ds \; .
\end{equation}
Thus, by using eq.(\ref{eq.p1s}), we can write
\begin{equation}
\frac{\Delta I(x)}{\tau} + I_0(x)= \int_0^\infty k(x,p) f_e(p) dp
\; ,
 \label{eq.diovertau}
 \end{equation}
where
\begin{equation}
k(x,p)=\int_{-\infty}^{+\infty} I_0(x e^{-s}) P_s(s,p) ds \; .
\end{equation}
Introducing the following definitions
\begin{eqnarray}
{\cal H}f_e(x)&\equiv &\int_0^\infty k(x,p) f_e(p) dp\\
\tilde{I}_\tau(x) & \equiv & \frac{I(x)}{\tau}+I_0(x),
\end{eqnarray}
the previous eq.\ref{eq.diovertau} writes as
\begin{equation}
\tilde{I}_\tau (x)={\cal H} f_e(x) \; .
\end{equation}
Its formal solution is given by
\begin{equation}
f_e(x)={\cal H}^{-1}\tilde{I}_\tau(x)
 \label{eq.formale}
\end{equation}
with
\begin{equation}
{\cal H}^{-1} {\cal H}={\cal I} \; ,
\end{equation}
where ${\cal I}$ is the identity operator.

\section{The SZE for a galaxy cluster with a radial temperature profile}

In order to apply the general formalism described in the previous
Section to a cluster with a spatially varying temperature and
density profiles (i.e., a cluster with a cool core or a
non-isothermal cluster), we must consider that:
\begin{itemize}
\item equations (\ref{spettro_ris})--(\ref{eq.pns}) still hold.
In particular, the expression for the optical depth remains the
same because it depends only on the gas density profile and not on
the cluster temperature;

\item the expression for the function $P(s)$, and its series expansion
in terms of powers of $\tau$, remains formally unchanged, because
it gives the probability that a CMB photon suffers a (logarithmic)
change in the frequency $s \equiv ln(\nu '/\nu)$ by traversing the
whole cluster;

\item the expression for the function $P_1(s)$
(i.e. eq. \ref{eq.p1s}) must be changed. In fact, the expression
given in eq.(\ref{eq.p1s}) assumes that the electron momentum
distribution function $f_e(p)$ is the same throughout the cluster
atmosphere. Such an assumption is not correct for a cluster with a
cool core (or a non-isothermal cluster) since its temperature, and
hence the electron momentum distribution, changes with the cluster
radius.
\end{itemize}

We must, therefore, find an appropriate way to calculate the
function $P_1(s)$ taking into account the fact that the CMB
photon, while traversing the cluster atmosphere, finds regions
with different temperature and density.
In order to calculate the redistribution function $P_1(s)$
appropriately, we can calculate -- along the line of sight $\ell$
-- the local electron momentum distributions weighted by its local
density.

Describing the thermal electron density as
\begin{equation}
n_e(r)=n_{e0}\cdot g_e(r) \; ,
\end{equation}
it is possible to derive the average density-weighted momentum
distribution along the line of sight $\ell$ as
\begin{equation}
\bar{f_e}(p)\equiv \frac{\int_\ell f_e(p;T_e(r)) g_e(r) d\ell}
{\int_\ell g_e(r) d\ell} \; ,
\label{febar}
\end{equation}
where
\begin{equation} \label{termica}
f_e(p;T_e)=\frac{\eta}{K_2(\eta)}p^2 \exp\left(-\eta
\sqrt{1+p^2}\right),
\end{equation}
with $\eta=(m_e c^2)/(k_B T_e)$ and $K_2 (\eta)$ being the
modified Bessel function of second kind  (see, e.g., Abramowitz \&
Stegun 1965), which ensures the correct normalization
\begin{equation}
\int_0^\infty f_e(p;T_e) dp =1 \; .
 \label{eq.normalizzazione}
\end{equation}
The function $\bar{f_e}$ also satisfies the normalization given in
eq.(\ref{eq.normalizzazione}), as can be verified by the following
equation
\begin{eqnarray}
\int_0^\infty \bar{f_e}(p) dp & = &  \frac{\int_0^\infty dp
\int_\ell f_e(p;T_e(r)) g_e(r) d\ell}{\int_\ell g_e(r) d\ell}= \nonumber\\
 & = & \frac{\int_\ell g_e(r) d\ell \int_0^\infty dp f_e(p;T_e(r))}
 {\int_\ell g_e(r) d\ell}=\nonumber \\
 & = & 1 \; \; ,
\end{eqnarray}
where we use eq. (\ref{eq.normalizzazione}).

We can, therefore, re-write eq.(\ref{eq.p1s}) for a cluster with a
generic temperature distribution by using the averaged electron
momentum distribution
\begin{equation}
P_1(s)=\int_0^\infty \bar{f_e}(p) P_s(s,p) dp \; ,
\label{eq.p1smedia}
\end{equation}
where $\bar{f_e}(p)$ is given in eq.(\ref{febar}). This function
allows us to calculate the frequency redistribution probability
for multiple scattering and, hence, the analytic expression of the
SZE up to the required approximation order by using eqs.
(\ref{spettro_ris})--(\ref{eq.pns}).

\section{Fitting procedure: the thermal SZE}

Eq.(\ref{eq.formale}) cannot be easily solved by analytically
methods. Therefore, we describe in this section a numerical
procedure by which it is possible to derive the cluster
temperature profile starting from observations of the SZE with
appropriate spatial and spectral resolution.\\
The quantitative analysis that we present here refers to the
specific case of the cluster A2199 which has a cool core.

The procedure that we follow consists of simulating the thermal
SZE observation of the cluster A2199 with parameters (density and
temperature) taken from the available X-ray observations, and then
extract the best fit values of the ICM temperature and optical
depth from the simulated SZE observations.
Note that in this section we work, for convenience, with the
first-order approximation to the thermal SZE, and in Sect.5 we
will discuss the consequences of using this approximation.
The temperature profile obtained from SZE measurements taken at
various cluster radii is then de-projected under the constraint
(prior) that the density profile derived from SZE observations is
consistent with the density profile obtained from X-ray
measurements: we use this procedure (which hence assumes the X-ray
density prior) in order to have a fast enough computing time that
allows us to explore the details of our study.
Note, however, that the density constraint is not necessary, in
principle, because the cluster density profile can be obtained,
self-consistently, from the optical depth profile measured from
SZE measurements. This fully consistent procedure requires,
however, much longer computing times since the number of free
parameters to be fitted to the observations is larger.

\begin{figure}[ht]
\begin{center}
 \epsfig{file=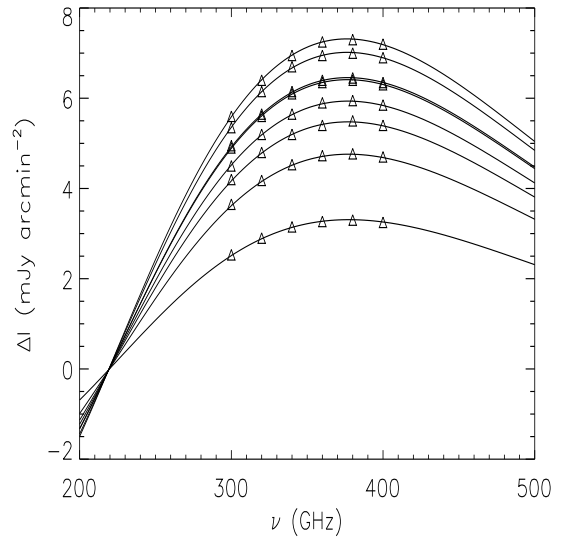,height=8.cm,width=8.cm,angle=0.0}
\end{center}
 \caption{\footnotesize{The SZE spectrum obtained for the cluster A2199
 at different projected radii is shown together with the frequency
 sampling derived, for each curve, with a 0.1\% uncertainty.
 From top to bottom curves refer to projected radii of
 0, 5, 10, 15, 20, 30, 50, 100 kpc. Note that spectra for 10 and 15
 kpc are almost superposed due to the fact that beyond this radial
 distance the IC gas temperature measured by Chandra decreases
 with increasing radius (Johnstone et al. 2002).}}
 \label{fig.campionamenti}
\end{figure}

\begin{figure}[ht]
\begin{center}
 \epsfig{file=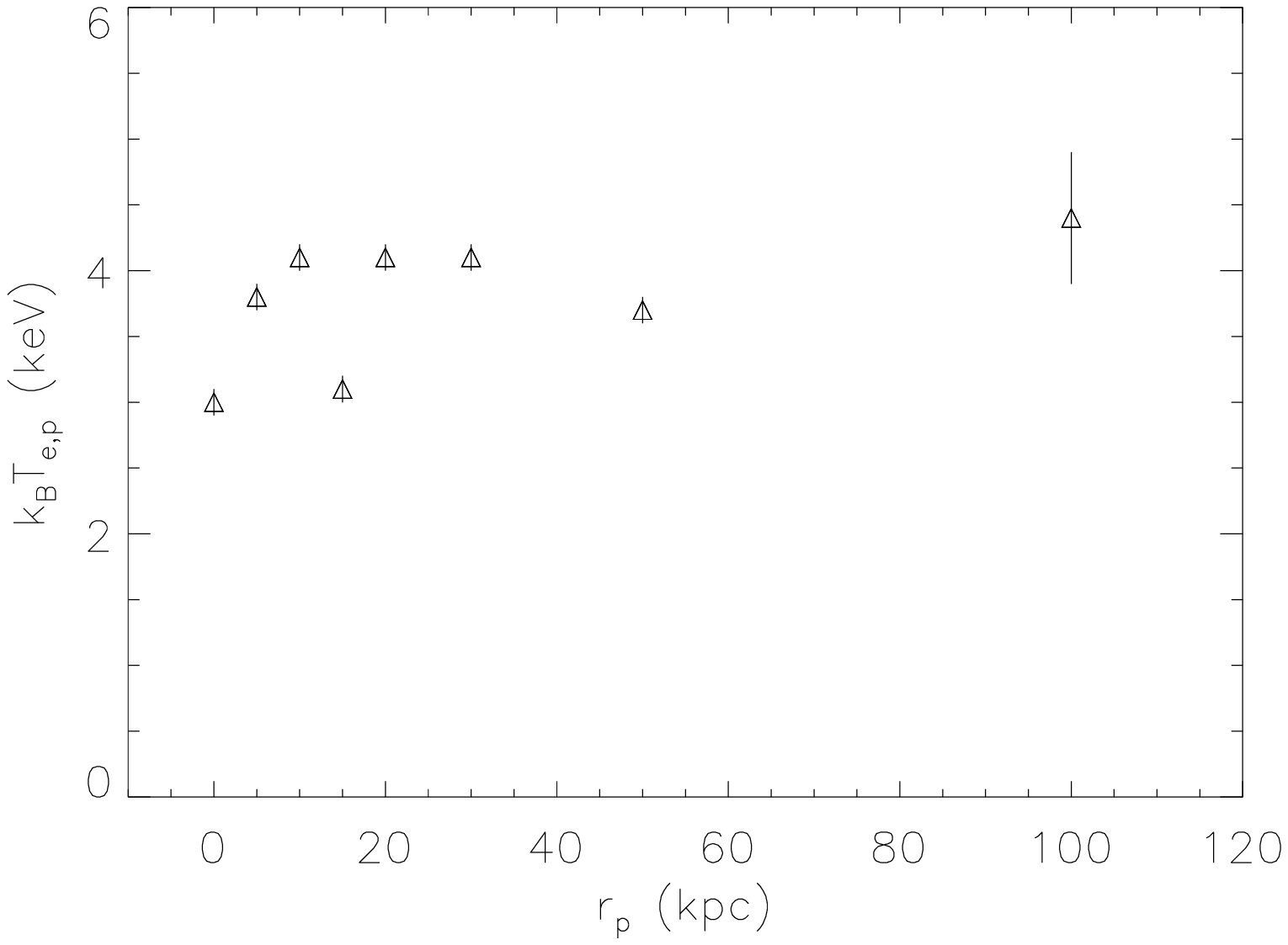,height=8.cm,width=8.cm,angle=0.0}
 \epsfig{file=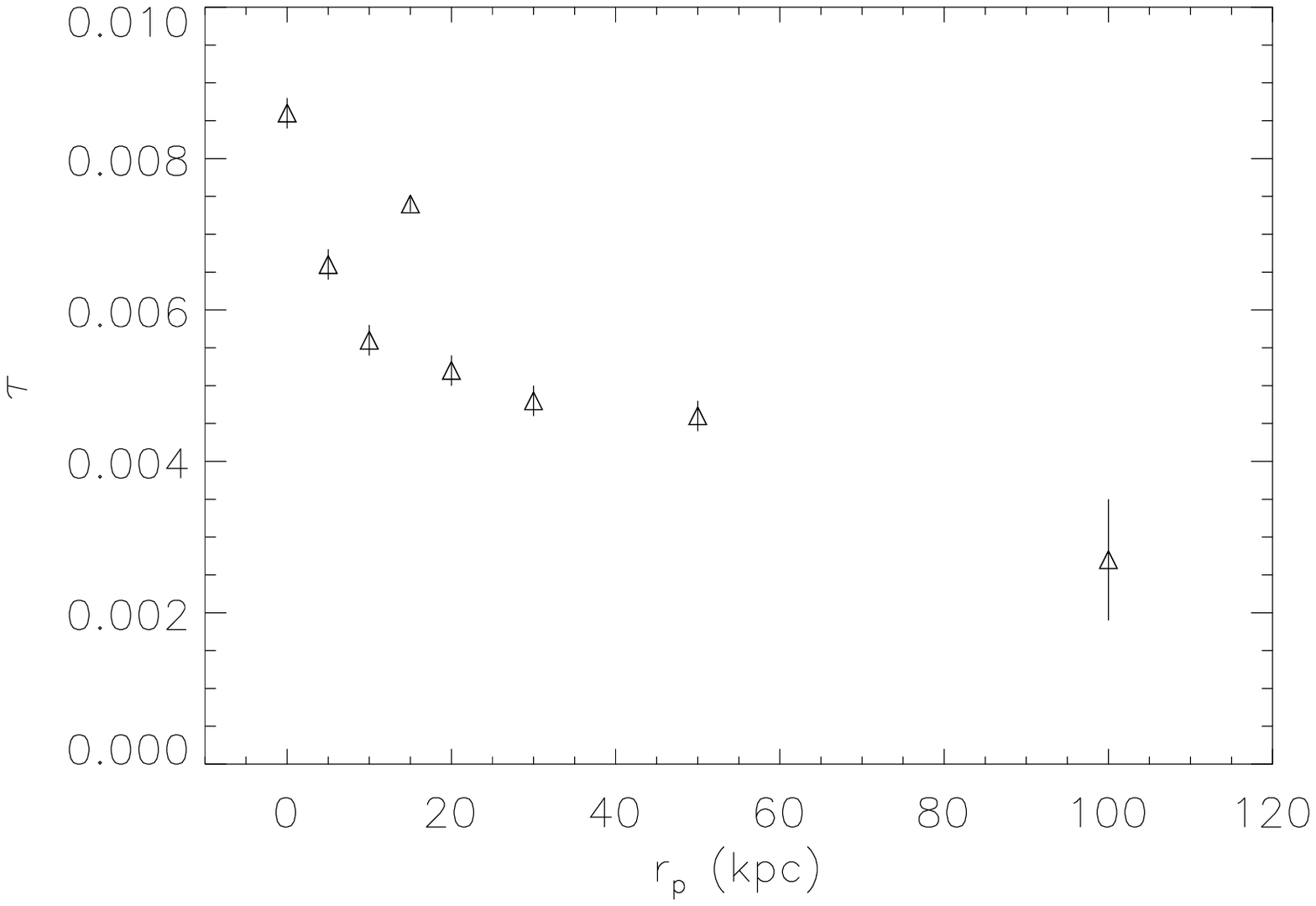,height=8.cm,width=8.cm,angle=0.0}
\end{center}
 \caption{\footnotesize{The radial profile of the projected
 temperature (top panel)
 and optical depth (bottom panel) as derived from the fit to the
 SZE spectra using data sampled in the frequency range 300--400 GHz
 with uncertainty of 0.1\% (see Fig.\ref{fig.campionamenti}).}}
 \label{fig.tproiett}
\end{figure}

\begin{figure}[ht]
\begin{center}
 \epsfig{file=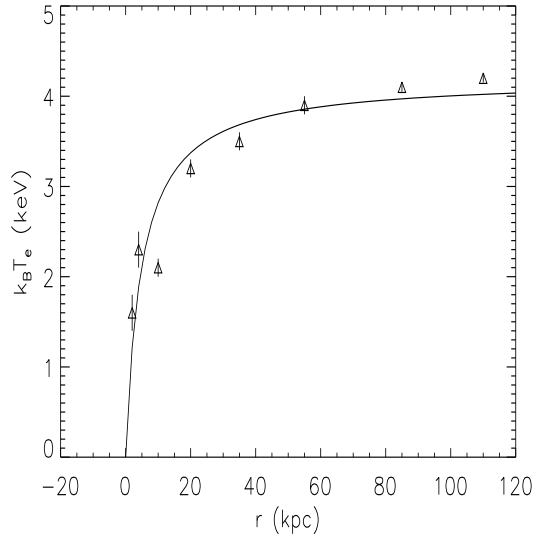,height=8.cm,width=8.cm,angle=0.0}
\end{center}
 \caption{\footnotesize{The deprojected temperature profile curve
 (solid) derived from SZE spectral data is compared with the
 deprojected temperature data as obtained from X-ray measurements
 of A2199 (Johnstone et al. 2002).}}
 \label{fig.temp.3d}
\end{figure}

For the reference case of A2199 that we want to discuss in this
section, our analysis goes thorugh the following steps:
i) we have taken X-ray information on A2199 from Chandra
observation (Johnstone et al. 2002);
ii) using the density and temperature profiles derived from X-ray
observations we  calculated (using the procedure described in
Sect.3 above) the SZE spectrum at eight different projected radii
$r_p$ from the cluster center, i.e., $r_p=$ 0, 5, 10, 15, 20, 30,
50 and 100 kpc;
iii) for each radial bin, we have sampled the SZE spectrum at six
different frequencies, i.e. $\nu=$ 300, 320, 340, 360, 380 and 400
GHz (see Fig. \ref{fig.campionamenti});
iv) for each radius, we fitted the six frequency SZE data with a
relativistic model of the thermal SZE leaving free two parameters,
i.e. $k_B T_e$ and $\tau$, and assuming for each experimental data
point an uncertainty of $0.1\%$. This fitting procedure
yields, for each projected radius, a value of the projected
temperature and optical depth (see Fig. \ref{fig.tproiett});
v) in order to deproject the fitted temperature profile, we have
assumed a 3-D theoretical temperature profile of the form:
\begin{equation}
T_e(r)=T_{int}+(T_{ext}-T_{int})\frac{(r/r_c)^\mu}{1+(r/r_c)^\mu}
\label{temp.3d}
\end{equation}
(see e.g. Piffaretti et al. 2005).
To decrease the number of free parameters in the fit we have
assumed the ICM density profile $n_e(r)$ as obtained from X-ray
data. In the fitting procedure, the projected temperature is
calculated by weighting the temperature profile with the gas
density and integrating along the los:
\begin{equation}
T_{e,p}(r_p)=\frac{2\int_0^R n_e(r)T_e(r)\frac{r}{\sqrt{r^2-r_p^2}}dr}
{2\int_0^R n_e(r)\frac{r}{\sqrt{r^2-r_p^2}}dr} \; ,
 \label{temp.weighted}
\end{equation}
where $R$ is the cluster radius (and it is $R=200$ kpc for the
case of A2199).
Assuming the value of $T_{ext}$ from X-ray observations, the
number of free parameters reduces to three. In this way it is
possible to fit the projected temperature as obtained from the
simulated SZE observations and then derive the temperature profile
parameter (in eq.\ref{temp.3d});
assuming $k_B T_{ext}=4.2$ keV for the case of A2199, the best fit
parameters are $k_B T_{int}=0.0\pm0.5$ keV, $r_c=4.9\pm0.6$ kpc and
$\mu=1.0\pm0.1$. The best-fit temperature profile curve is shown
in Fig.\ref{fig.temp.3d} and it is compared to the data on the
de-projected temperature as derived from X-ray observations.

\begin{figure}[ht]
\begin{center}
 \epsfig{file=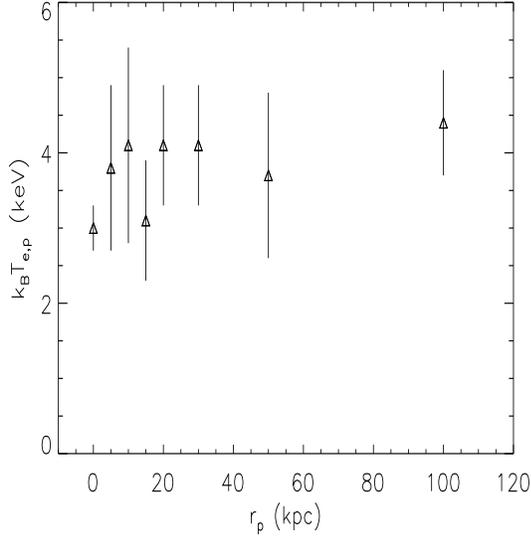,height=8.cm,width=8.cm,angle=0.0}
\end{center}
 \caption{\footnotesize{The projected temperature profile obtained
 from the fit to SZE spectra using data points sampled in the 300--400 GHz
 range and assuming an uncertainty of 1\%.}}
 \label{fig.tproiettbis}
\end{figure}

The same procedure has been also repeated assuming that the
SZE observation sampled from Fig. \ref{fig.campionamenti} has an
uncertainty of $1\%$. The radial profile of the projected
temperature obtained in this case is shown in Fig.
\ref{fig.tproiettbis}. In this case, the de-projection of the
temperature profile using the template given in eq.
(\ref{temp.3d}) does not provide, however, an acceptable fit. For
this reason we conclude that an uncertainty of order of $1\%$ in
the SZE data is not sufficient to derive a detailed
profile of the ICM temperature in clusters like A2199.

\begin{figure}[ht]
\begin{center}
 \epsfig{file=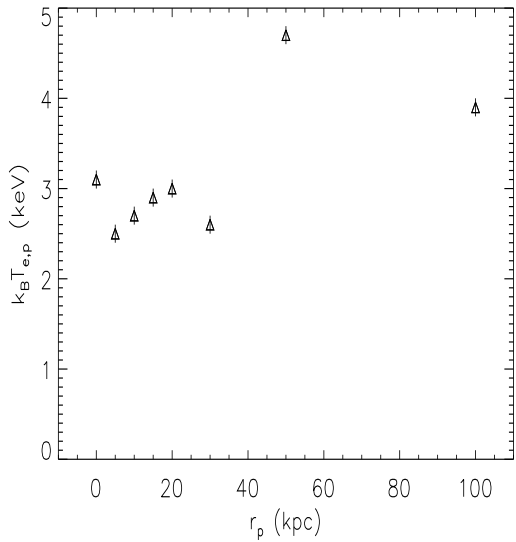,height=8.cm,width=8.cm,angle=0.0}
\end{center}
 \caption{\footnotesize{The projected temperature radial profile
 obtained from the fit to the SZE spectral data by using data points sampled in
 100--200 GHz range and assuming uncertainties of 0.1\%.}}
 \label{fig.tproiett2}
\end{figure}

The whole procedure has been repeated sampling the SZE
observations in the low-frequency range 100--200 GHz
($\nu=100$, 120, 140, 160, 180, 200 GHz).
In such a case, we find that, even assuming uncertainties of 0.1\%,
it is not possible to reproduce in a
satisfactory way the radial profile of the projected temperature
as derived from the fit by using the model in eq.(\ref{temp.3d})
(see Fig.\ref{fig.tproiett2}). This is because the shape of the
SZE spectrum at low frequencies is less sensitive to the cluster
temperature than the shape of the high-frequency part of the SZE
spectrum. Therefore, in order to obtain detailed information on
the cluster temperature profile one must use high-frequency
spectral observations in the range 300--400 GHz. This is a crucial
requirement to plan an observational strategy of SZE observations
with the goal of deriving strong constraints to the fundamental
parameters of the cluster.

\section{Fitting procedure: the non-thermal SZE}

Here we consider the effects of the variation of the parameters of
the non-thermal electron spectrum on the relative non-thermal SZE.

\begin{figure}[ht]
\begin{center}
 \epsfig{file=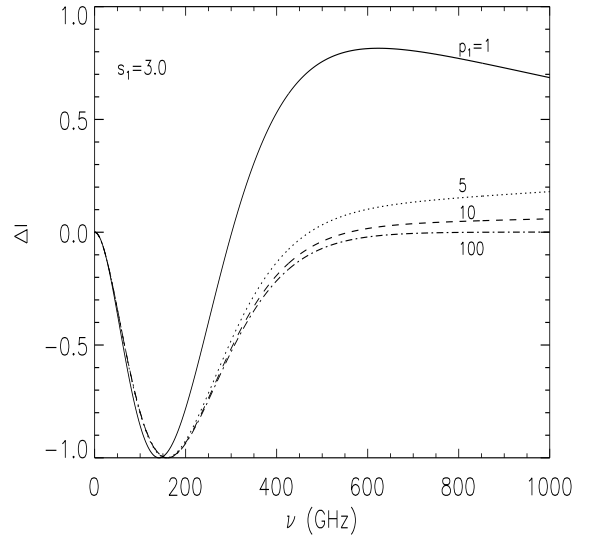,height=8.cm,width=8.cm,angle=0.0}
\end{center}
 \caption{\footnotesize{The non-thermal SZE spectrum, normalized to its
 minimum amplitude is shown as a function of the minimum momentum
 $p_1$ of the electron spectrum. We assume here a single power-law spectrum with $s_1=3.0$.}}
 \label{fig.SZnt_p1}
\end{figure}

First, we study the spectral shape of the non-thermal SZE as a
function of the electrons spectral parameters. In this case, we
use an arbitrary normalization in which the amplitude of the SZE
at its minimum is set equal to $-1$, and we vary the
parameters of the electrons spectrum. Fig.\ref{fig.SZnt_p1} shows
the SZE spectrum for a single power-law spectrum
\begin{equation}
f_e(p)\propto p^{-s_1} \; ; \;\;\;\; p\geq p_1
\end{equation}
(here $p=\beta\gamma$ is the normalized momentum), with $s_1=3.0$,
as a function of the minimum momentum $p_1$. It is clear that the
frequency position of the minimum of the SZE is different in the
case $p_1=1$, while for $p_1\geq5$ the position of the minimum
does not change for different values of $p_1$. The shape of the
spectrum at higher frequencies ($\nu \geq 300$ GHz) is, instead,
more sensitive to the value of $p_1$.

\begin{figure}[ht]
\begin{center}
 \epsfig{file=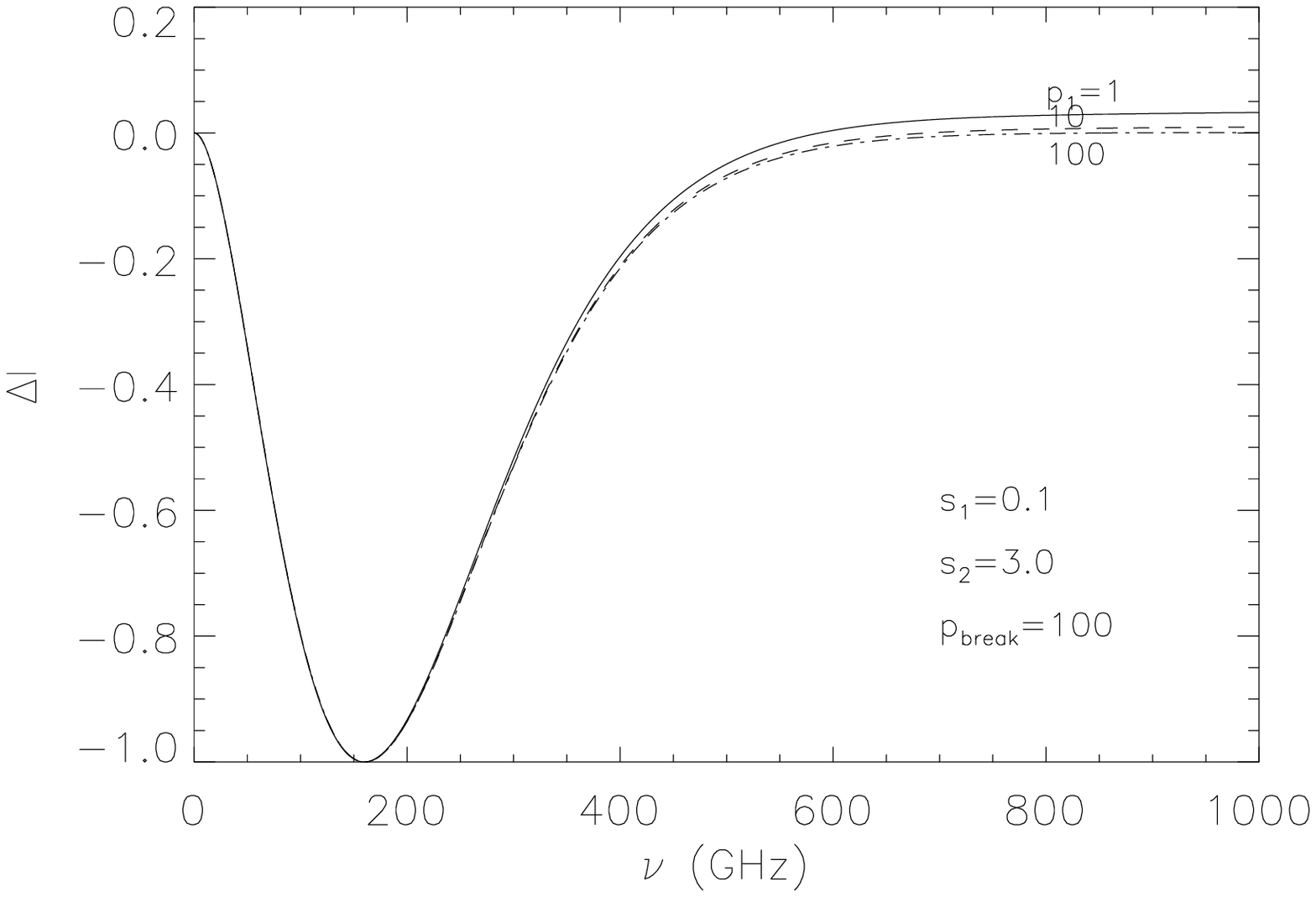,height=8.cm,width=8.cm,angle=0.0}
\end{center}
 \caption{\footnotesize{The non-thermal SZE spectrum, normalized to its
 minimum amplitude is shown as a function of the minimum momentum
 $p_1$ of the electron spectrum. We assume here a double power-law spectrum with
 $s_1=0.1$, $s=3.0$ and $p_{break}=100$.}}
 \label{fig.SZnt_p1_2lp}
\end{figure}

We show in Fig.\ref{fig.SZnt_p1_2lp} the case of a double
power-law spectrum
\begin{equation}
f_e(p)\propto\left\{ \begin{array}{cc}
p^{-s_1} & p_1\leq p \leq p_{break} \\
p_{break}^{-s_1}(p/p_{break})^{-s_2} & p>p_{break}
\end{array} \right.
\label{eq.spettro_doppialp}
\end{equation}
with $s_1=0.1$, $p_{break} =100$ and $s_2=3.0$. In this case the
position of the minimum and the shape of the spectrum at
frequencies $\nu\leq400$ GHz, depends weakly from the value of
$p_1$, while a stronger dependence from $p_1$ is present at higher
frequencies (this dependence is anyway weaker than for the single
power-law case because of the lower power at the low-$p$ end of
the electron spectrum).

\begin{figure}[ht]
\begin{center}
 \epsfig{file=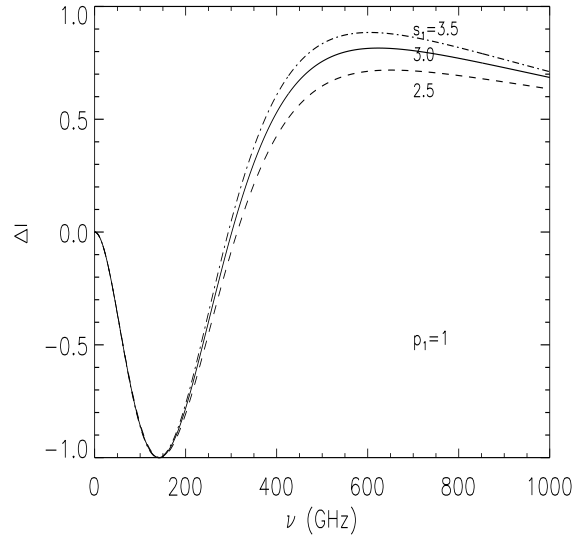,height=8.cm,width=8.cm,angle=0.0}
\end{center}
 \caption{\footnotesize{The non-thermal SZE spectrum, normalized to its
 minimum amplitude is shown as a function of the spectral index $s_1$
 of the electron spectrum. We assume here a single power-law spectrum with $p_1=1$.}}
 \label{fig.SZnt_s}
\end{figure}

\begin{figure}[ht]
\begin{center}
 \epsfig{file=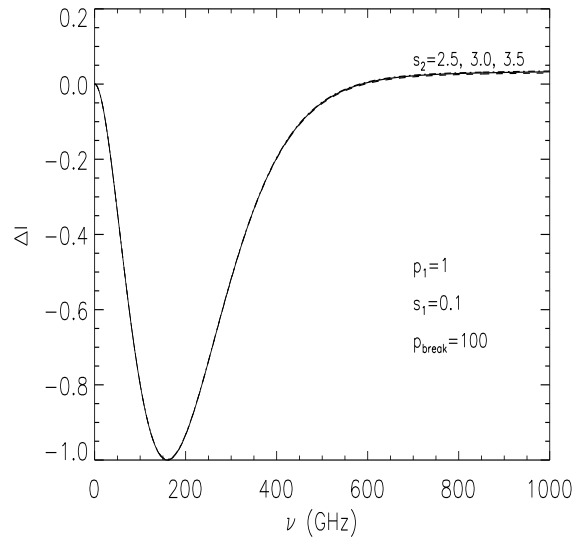,height=8.cm,width=8.cm,angle=0.0}
\end{center}
 \caption{\footnotesize{The non-thermal SZE spectrum, normalized to its
 minimum amplitude is shown as a function of the spectral index $s_2$ of the electron spectrum.
 We assume here a double power-law spectrum with $p_1=1$, $s_1=0.1$ and $p_{break}=100$.}}
 \label{fig.SZnt_s_2lp}
\end{figure}

Fig.\ref{fig.SZnt_s} shows the changes of the SZE spectrum in the
case of a single power-law with $p_1=1$ as a function of the
spectral index $s_1$: also in this case the shape of the SZE
spectrum is more sensitive to the shape of the electron spectrum
at frequency $\nu\geq400$ GHz.
Fig.\ref{fig.SZnt_s_2lp} show the variation of the SZE spectrum as
a function of the spectral index $s_2$ for a double power-law
spectrum: in this case the dependence from the electron spectrum
index is quite weak for the considered frequency range.

After having shown the dependence of the SZE spectrum on the
electron spectrum parameters, we discuss now the procedures
through which we can recover the physical parameters of the
non-thermal electron spectrum from the SZE observations.
We consider here two cases: i) the case of a pure non-thermal
electron population with a power-law spectrum (a situation
applicable to radio galaxy lobes, see e.g. Colafrancesco 2008b);
ii) the case of a combination of thermal plus non-thermal electron
population with a double power-law spectrum (a situation
applicable to galaxy clusters with non-thermal phenomena, see e.g.
Colafrancesco et al. 2003).

\subsection{The case of a single power-law spectrum}

We describe here the procedure through which it is possible to
derive the parameters of an electron population with single
power-law spectrum, assuming that this is the dominant electron
population for the SZE production.

Following the same procedure previously outlined for the thermal
SZE in Sect.4, we have sampled the non-thermal SZE produced at
different projected radii and at six frequencies in the range 300
-- 400 GHz where, as previously noticed, the shape of the SZE is
maximally sensitive to the physical parameters of the electron
spectrum.
The emission region has been assumed, for simplicity, to have
spherical symmetry with a radius of 50 kpc with an electron
spectrum
\begin{equation}
N_e(p,r) = k_0 p^{-s_1} \cdot g_e(r) \;\;\;\;\; p\geq p_1 \; \; .
\label{eq.elent}
\end{equation}
Here we assume $s_1=3.0$, $p_1=3$, $k_0=10^{-2}$ cm$^{-3}$ and a
radial distribution of the electron population given by
\begin{equation}
g_e(r)=\left[ 1+ \left(\frac{r}{r_c}\right)^2 \right]^{-q_e}
\label{eq.elespaz}
\end{equation}
with $r_c=10$ kpc and $q_e=0.5$.
%

\begin{figure}[ht]
\begin{center}
 \epsfig{file=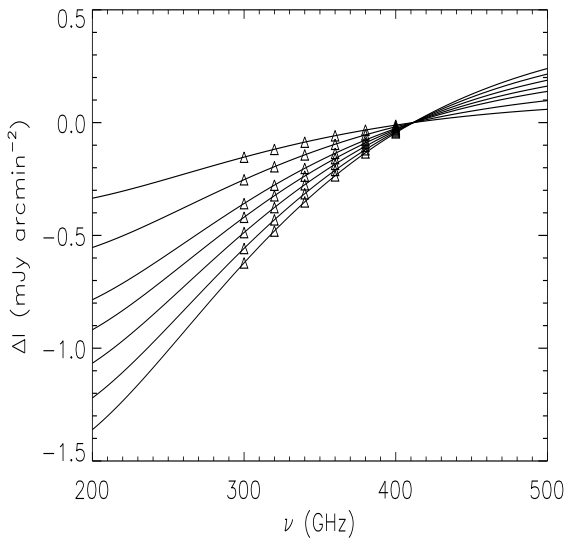,height=8.cm,width=8.cm,angle=0.0}
\end{center}
 \caption{\footnotesize{The non-thermal SZE spectrum evaluated at
 various projected radii and frequency sampling (see text for
 details) with  data uncertainty of 0.1\%.
 Curves are, from bottom to top, for projected radii of
 0, 5, 10, 15, 20, 30, 40 kpc.}}
 \label{fig.campionamenti_nt}
\end{figure}

\begin{figure}[ht]
\begin{center}
 \epsfig{file=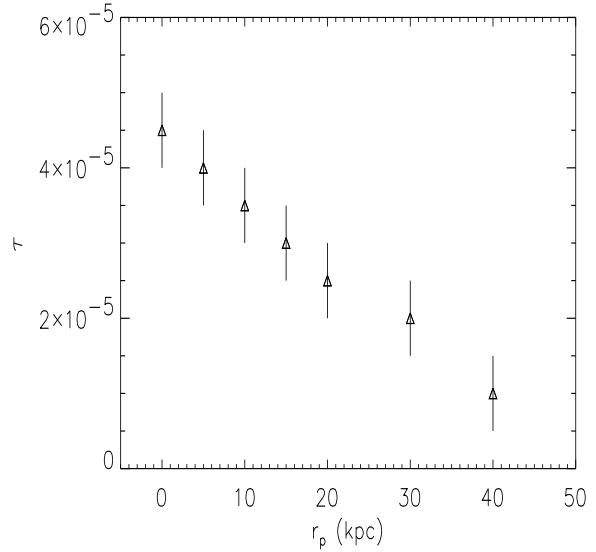,height=8.cm,width=8.cm,angle=0.0}
\end{center}
 \caption{\footnotesize{The radial profile of the optical depth as obtained
 from the fit to the SZE spectra b using data sampled in the range
 300--400 GHz with uncertainty of 0.1\%
(see Fig.\ref{fig.campionamenti_nt}).}}
 \label{fig.tau_nt}
\end{figure}

In the fitting procedure we discuss here, we fix the spectral index of the
electron spectrum $s_1$, because it can be derived from radio
observations, and we fit the simulated observations with two free
parameters: the optical depth $\tau$ and the minimum momentum
$p_1$.
We note that the assumption of the parameter $s_1$
from radio data is not strictly necessary,
because, in the presence of adequate coverage in frequency, it can be
derived from the SZE data alone; this assumption is made for the
sole purpose of reducing the time of calculation.\\
The SZE intensity at different radii and the relative sampling are
shown in Fig.\ref{fig.campionamenti_nt}.
The behaviour of the optical depth at different projected radii,
as derived by the fitting procedure, is shown in Fig.
\ref{fig.tau_nt}. Our analysis recovers the values $p_1=3.1\pm1.0$
at all the considered projected radii.
In order to deproject the optical depth radial distribution and
derive the radial profile of the electron density it is sufficient
to fit the optical depth profile with a density profile as in eq.
(\ref{eq.elespaz}) and integrate along the line of sight. The
deprojection procedure yields  $k_0=(9.0\pm3.5)\times10^{-3}$
cm$^{-3}$, $r_c=7.2\pm6.0$ kpc and $q_e=0.43\pm0.21$; these values
are consistent, within the errors, with the input values that we
used in the simulations.

We stress that the previous analysis of the SZE spectral
measurements allows to derive directly the density profile and the
spectral shape of the electron distribution in the radio lobes.
Therefore, the combination of SZE spectral measurements and
measurement of the low-frequency radio emission allows to break
the degeneracy existing in synchrotron emission between the
electron density and the magnetic field and eventually yield a
measure of the magnetic field in radio lobes (see also
Colafrancesco 2008b).

\subsection{The case of thermal plus non-thermal populations}

Here we consider a combination of a thermal electron distribution
that fits the A2199 data and an additional non-thermal electron
spectrum with a double power-law spectrum, as in
eq.(\ref{eq.spettro_doppialp}), with parameters $s_1=0.1$,
$s_2=3.0$, $p_1=1$ and $p_{break}=100$.

A non-thermal electron spectrum similar to a double power-law is
expected in galaxy clusters because the non-thermal electrons of
relatively low energy loose energy through Coulomb interactions
with the thermal IC gas, and these energy losses produce a
flattening of the electron spectrum at low ($\simlt 150$ MeV)
energies.\\
To study this case, we do not follow the same strategy described
in the previous Sections: specifically, we do not perform a fit to
the sampled points because this procedure should involve too many
free parameters (i.e., those of both the thermal and the
non-thermal electron populations). Instead, we focus on the
spectral regions where the non-thermal effect is expected to be
more relevant, and therefore it can be measured. For this purpose,
we compare the non-thermal SZE spectrum with the thermal SZE
spectrum (calculated both at first order in $\tau$ and with second
order corrections, see Sect.2 and Colafrancesco et al. 2003 for
more details), and with the spectrum of a possible SZ kinematic
effect. The spectral regions where the non-thermal SZE could be
better estimated are the frequency range around the minimum of the
SZE (at $\nu \sim 150 $ GHz), because all the amplitudes are
basically proportional to the optical depth of the relative
electronic population, and the frequency range around the
crossover frequency of the thermal SZE (at $\nu \sim 220 $ GHz),
because the thermal SZE is by definition null at such frequency
leaving hence visible the (negative in sign) non-thermal SZE.

\begin{figure}[ht]
\begin{center}
 \epsfig{file=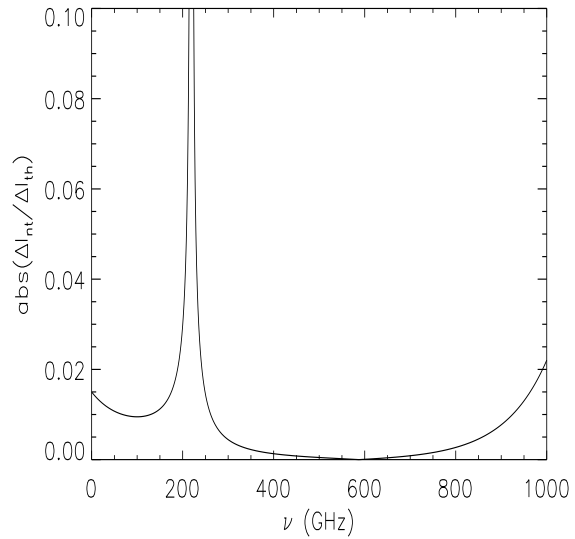,height=8.cm,width=8.cm,angle=0.0}
\end{center}
 \caption{\footnotesize{The frequency dependence of the ratio of the non-thermal SZE
 (we assume a double power-law spectrum with
 $s_1=0.1$, $s_2=3.0$, $p_1=1$ and $p_{break}=100$) to the thermal
 SZE at the center of A2199. We assume a pressure balance
 $P_{nt} \sim P_{th}$ at the cluster center.}}
 \label{fig.errore_nt}
\end{figure}

\begin{figure}[ht]
\begin{center}
 \epsfig{file=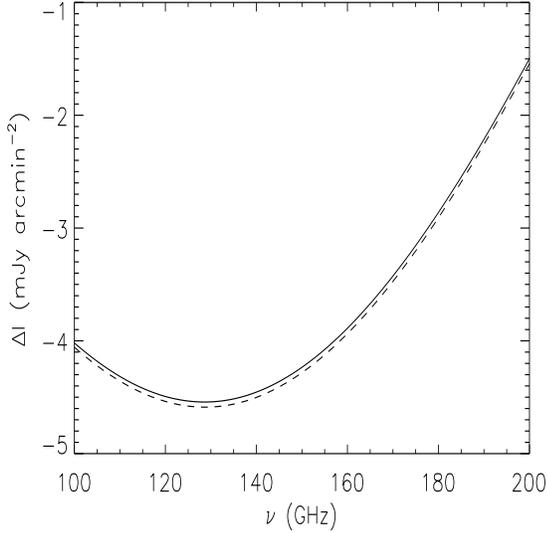,height=8.cm,width=8.cm,angle=0.0}
 \epsfig{file=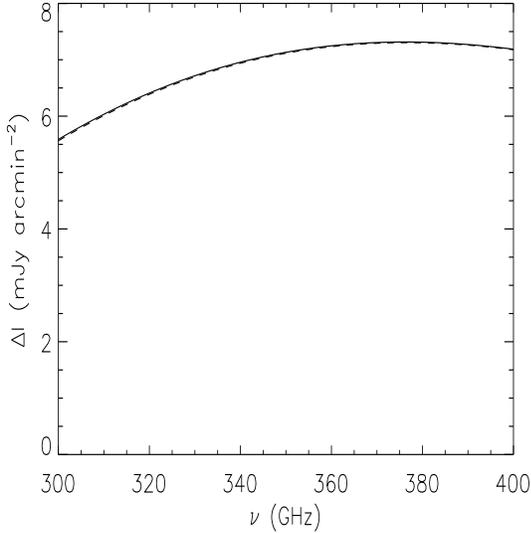,height=8.cm,width=8.cm,angle=0.0}
\end{center}
 \caption{\footnotesize{The thermal (solid curve) and non-thermal (dashed curve)
 SZE evaluated at the center of A2199. The non-thermal electron
 population has a double power-law spectrum with parameters
$s_1=0.1$, $s_2=3.0$, $p_1=1$ and $p_{break}=100$ and a pressure
$P_{nt}\sim P_{th}$ at the cluster center. The spectra are shown
in the frequency range 100--200 GHz (upper panel) and 300--400 GHz
(lower panel).}}
 \label{fig.spettrisz_th+nt}
\end{figure}

We note that an upper limit to the non-thermal electron density
can be derived, after fixing their spectral shape, by requiring
that the non-thermal pressure does not exceeds the thermal one
(see discussion in Colafrancesco et al. 2003).
Fig.\ref{fig.errore_nt} shows the ratio between the non-thermal
SZE and the thermal SZE in the case $P_{nt}\sim P_{th}$ at the
cluster center: this case can be considered as an upper limit to
the contribution of the non-thermal component to the total SZE.
Fig. \ref{fig.spettrisz_th+nt} shows the spectrum of the thermal
and non-thermal SZE in the frequency ranges 100--200 GHz and
300--400 GHz.
The ratio of the non-thermal to the thermal SZE is large only in
the region around $\nu\sim220$ GHz where the thermal SZE has its
crossover and where, consequently, the total SZE is dominated by
the non-thermal component. For $\nu<200$ GHz the non-thermal SZE
is $\sim 1\%$ of the thermal one, while for $\nu>300$ GHz the
ratio becomes very low, of order of $\sim 0.1-0.5\%$, or even
less.
Two conclusions can be derived from these results:
1) In the frequency range 300--400 GHz, that is optimal to derive
precise information on the thermal SZE, the non-thermal component
does not provide a contribution that jeopardizes the precision of
the measurements.
2) At lower frequencies, around 200 GHz, the contribution of the
non-thermal component is more relevant; however, to determine such
a component, it is necessary to separate it from the thermal one.

A possible observational strategy to separate the non-thermal
effect from the thermal one could consist in measuring the thermal
SZE in the frequency range 300--400 GHz and deriving the IC gas
physical parameters, as seen in Sect.4. Then, once the thermal SZE
parameters are fixed, it will be possible to measure the SZE in
the frequency range 100--200 GHz and derive the parameters of the
non-thermal population that, in this frequency range, yields a
contribution to the total SZE of order of  $\sim 1\%$ (see figures
\ref{fig.errore_nt} and \ref{fig.spettrisz_th+nt}): such SZE
amplitude is measurable if measurements have an uncertainty of
order of  $\sim 0.1\%$.

\begin{figure}[ht]
\begin{center}
 \epsfig{file=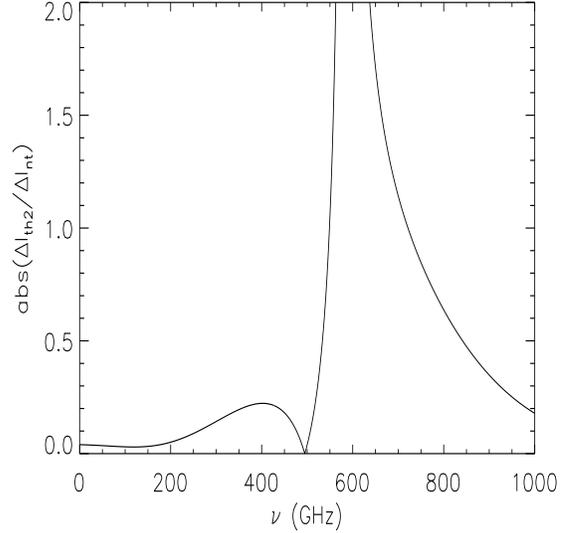,height=8.cm,width=8.cm,angle=0.0}
 \epsfig{file=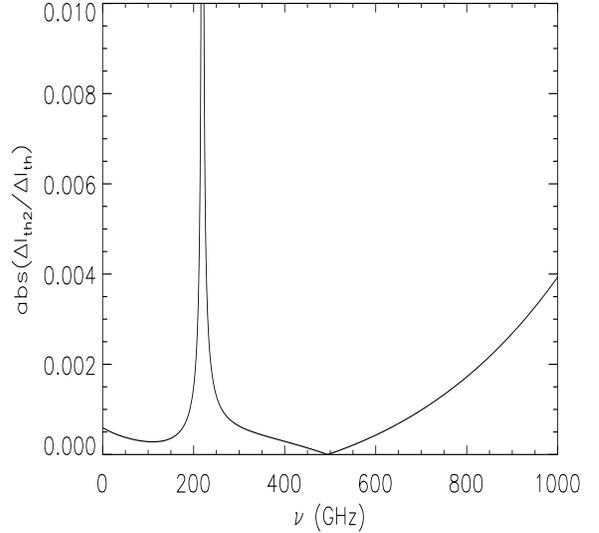,height=8.cm,width=8.cm,angle=0.0}
\end{center}
 \caption{\footnotesize{Upper panel: the frequency dependence of
 the absolute value of the ratio between
 the second order correction to the thermal SZE at the center of
 A2199 and the non-thermal SZE of an electron population with
 double power-law spectrum with parameters
$s_1=0.1$, $s_2=3.0$, $p_1=1$, $p_{break}=100$ and with pressure
$P_{nt}\sim P_{th}$. Lower panel: the frequency dependence of the
ratio between the second order correction and the first order
description of the thermal SZE.}}
 \label{fig.errore_th2}
\end{figure}

For a precise determination of the cluster parameters, we must
take into account the contribution of the thermal SZE evaluated at
second order in $\tau$; this can be calculated in details once the
main parameters of the IC gas are known (see Colafrancesco et al.
2003 for details). The modifications induced by the second order
effect, in comparison to the non-thermal SZE and to the
first-order thermal SZE, are shown in Fig. \ref{fig.errore_th2}.
For the frequency range 300--400 GHz, that is the optimal
frequency window to extract the thermal SZE parameters, the
second-order correction to the thermal SZE is less than 30\% of
the non-thermal SZE and it is less than 0.1\% of the first-order
thermal SZE. Hence, we can conclude that the total SZE can be
adequately approximated, in this frequency range, by the
first-order thermal SZE only. Moreover, in the frequency range
100--200 GHz, where the non-thermal effect could be measured, the
second order correction to thermal effect is $\sim5\%$ of the
non-thermal effect, so that the second order correction can be
neglected also for this purpose.

\begin{figure}[ht]
\begin{center}
 \epsfig{file=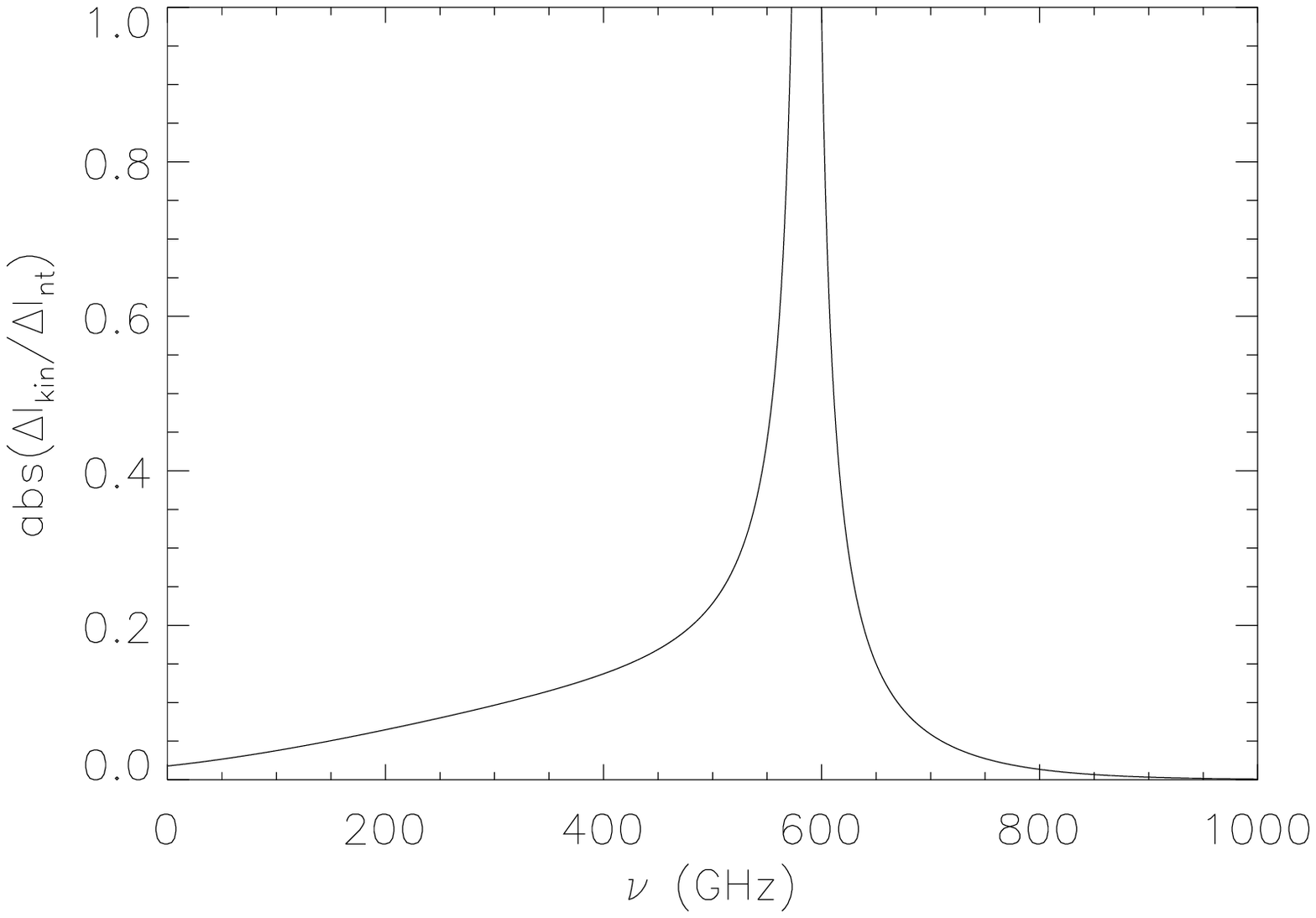,height=8.cm,width=8.cm,angle=0.0}
 \epsfig{file=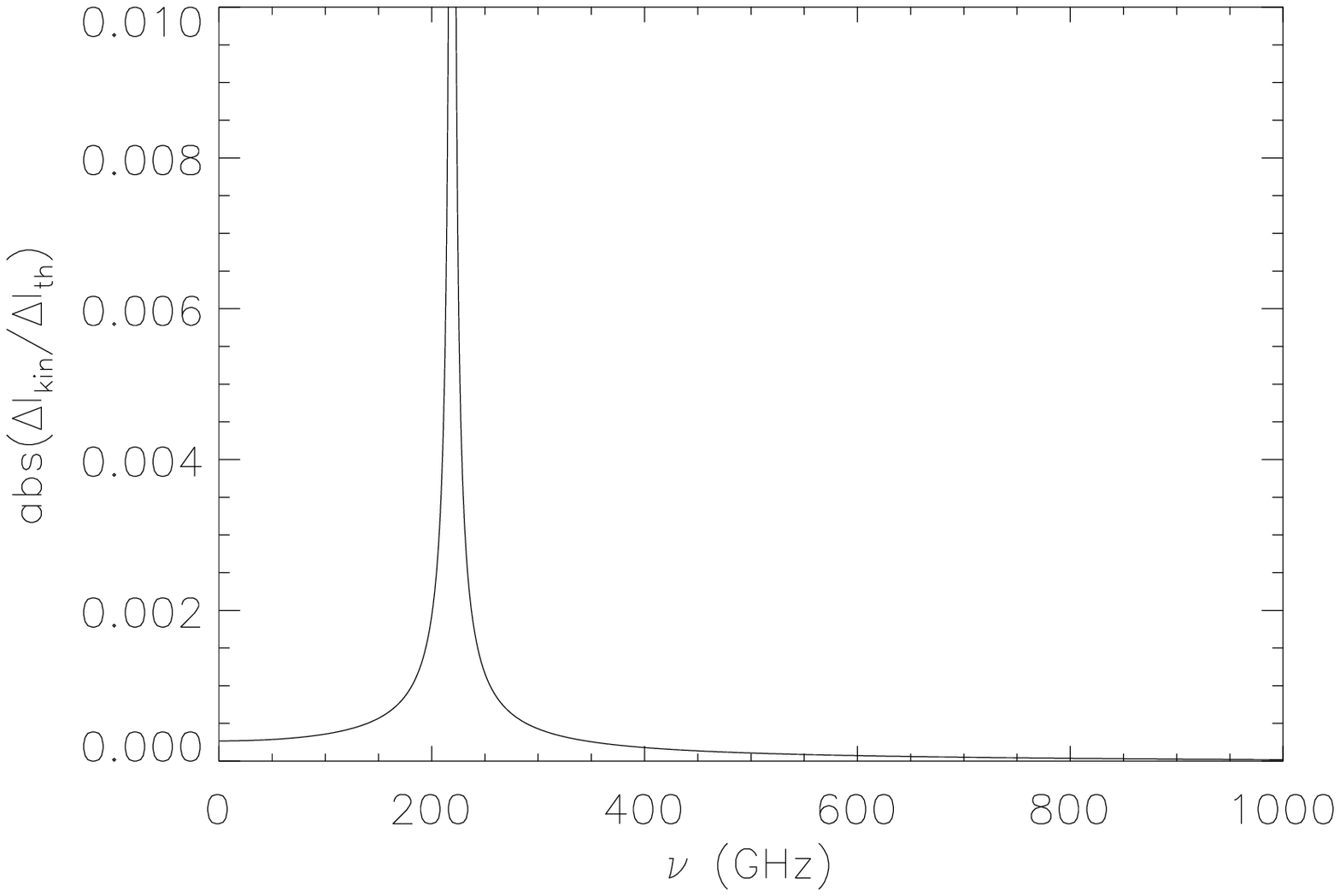,height=8.cm,width=8.cm,angle=0.0}
\end{center}
 \caption{\footnotesize{Upper panel: the frequency dependence of
 the absolute value of the
 ratio between the kinematic SZE for a value of the peculiar
 velocity of $v_p=1000$ km/s and the non-thermal SZE for an
 electron population with the same spectrum as in Fig. \ref{fig.errore_th2}.
 Lower panel: the frequency dependence of the absolute value of the ratio between
 the kinematic SZE and the thermal SZE evaluated at first order.}}
 \label{fig.errore_cin}
\end{figure}

We consider now also the possible contribution of a kinematic SZE
whose amplitude is maximal at frequency around the crossover
frequency of the thermal SZE, i.e. at $\nu\sim220$ GHz. Fig.
\ref{fig.errore_cin} shows the ratio of the kinematical to the
non-thermal SZE (upper panel) and the ratio of the kinematical to
the thermal SZE (lower panel) for a receding cluster with peculiar
velocity of 1000 km/s. The result is that in the frequency range
300--400 GHz also the kinematical contribution is a small fraction
($\sim 20\%$) of the non-thermal SZE, and it is less than 0.1\% of
the thermal one. Therefore, also the kinematical SZE can be
neglected in this frequency range.\\
Fig. \ref{fig.errore_cin} also shows that in the range 100--200
GHz the kinematical SZE is less than 10\% of the non-thermal SZE.
However, if the non-thermal electron population have a pressure
$P_{nt}\sim 0.1 P_{th}$, then the non-thermal and kinematical SZE
would become comparable (under the assumption of having $V_p =
1000$ km/s); in such a case, however, these signals would be
indistinguishable from the thermal one (see for reference,  Figs.
\ref{fig.errore_nt} and \ref{fig.spettrisz_th+nt} upper panel, in
which we show the case $P_{nt}\sim P_{th}$).

We can summarize the results obtained:\\
- in the frequency range 300--400 GHz, the first order thermal SZE
is a good approximation of the total SZE (all the corrections are
$\leq 0.1\%$).\\
- in the frequency range 100--200 GHz, if the non-thermal
contribution is strong, it dominates over the kinematic one. If,
instead, the two SZE contributions are comparable, then they
cannot be easily disentangled from the thermal one. Hence, if
deviations from expected thermal SZE are measured in this
frequency range, they should be of non-thermal origin.\\
- only at $\nu\sim 220$ GHz (i.e. the crossover frequency of
thermal SZE) the kinematic SZE, the non-thermal one and the second
order thermal correction are important with respect to the first
order thermal effect. To disentangle between the three different
contributions it is important to put constraints on the thermal
and the non-thermal parameters by performing precise observations
in frequency ranges 300--400 and 100--200 GHz, respectively.

\section{Applications to specific clusters: Perseus and Ophiuchus}

We present in this section the results of our studies for two
nearby and extended clusters: Perseus, which has a cool core, and
Ophiuchus, which is approximately isothermal. As we discussed in the previous
sections, the optimal strategy to extract the physical parameters
of the cluster is to consider the frequency range 300--400 GHz; in
this range, we can safely consider only the first order thermal
SZE in our analysis, because higher-order corrections are
negligible.

\subsection{The Perseus cluster}

The Perseus cluster is at a redshift $z=0.0179$ or at a distance
of 77.7 Mpc, at which 1 arcmin corresponds to 21.8 kpc. Therefore,
a study of the spatially resolved SZE allows to have a spatial
sampling of the temperature profile better than in the case of
A2199 previously discussed. In the following, we repeat the
previous analysis for this cluster.

Churazov et al. (2003) derived an analytical approximation for the
radial profile of the IC gas density and of the gas temperature
that fits the cluster data for $r>10$ kpc, while at shorter
distances the presence of the central dominating galaxy NGC1275
makes the IC gas radial profile more uncertain. Here, for sake of
illustration, we use the analytical fitting formulae given by
Churazov et al. (2003) and we estrapolate them down to the cluster
center, e.g. ignoring, in this way, the presence of the central
galaxy.

\begin{figure}[ht]
\begin{center}
 \epsfig{file=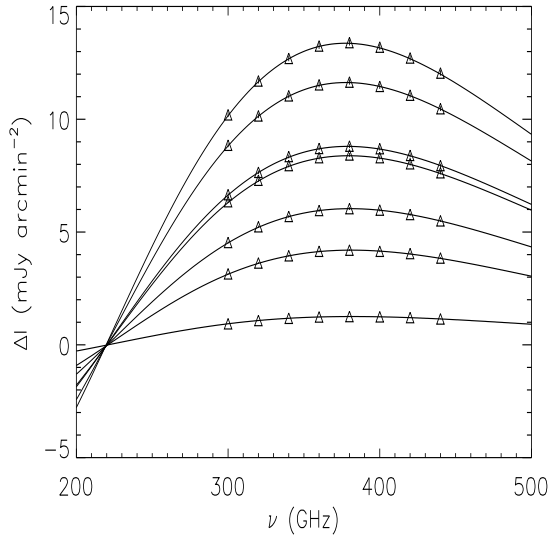,height=8.cm,width=8.cm,angle=0.0}
\end{center}
 \caption{\footnotesize{The thermal SZE spectra of the Perseus cluster
 evaluated at different projected radii of 0, 20, 50, 75, 100, 150 e 200
 kpc (from top to bottom) are shown in the frequency range 200--500 GHz.
 For each curve the SZE sampling in the frequency range 300--450 GHz
 has an uncertainty of 1\%.}}
 \label{fig.campionamenti_pers}
\end{figure}

\begin{figure}[ht]
\begin{center}
 \epsfig{file=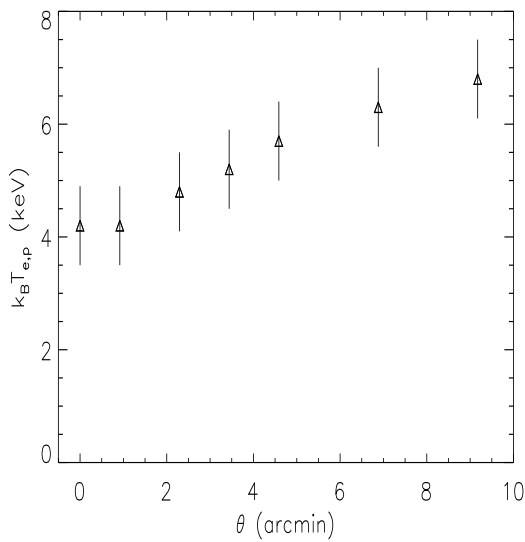,height=8.cm,width=8.cm,angle=0.0}
\end{center}
 \caption{\footnotesize{The radial profile of the projected temperature,
 as derived from the fit to the SZE spectra in Perseus in the
 frequency range 300--440 GHz with uncertainties of 1\% (see Fig.
\ref{fig.campionamenti_pers}), is shown as a function of the
projected radius (in arcmin).}}
 \label{fig.tproiett_pers}
\end{figure}

We have hence simulated the thermal SZE to first order expected
from Perseus at seven projected radii from the cluster center: 0,
20, 50, 75, 100, 150 and 200 kpc (corresponding to angular sizes
of 0, 0.9, 2.3, 3.4, 4.7, 6.9 and 9.2 arcmin). We have then
sampled each one of the SZE spectra at these radii for eight
different frequencies: 300, 320, 340, 360, 380, 400, 420 and 440
GHz (see Fig. \ref{fig.campionamenti_pers}).
We have extended the frequency coverage in order to better sample
the spectral region of the SZE where relativistic effects are relevant.
For each projected
radius we have finally fitted, using the gas temperature and
optical depth as free parameters, the SZE simulated signal by
using the expression of the first-order thermal SZE and assuming
an error of 1\% in the experimental data.
The results of the fit for the projected temperature are shown in
Fig. \ref{fig.tproiett_pers}.
\begin{figure}[ht]
\begin{center}
 \epsfig{file=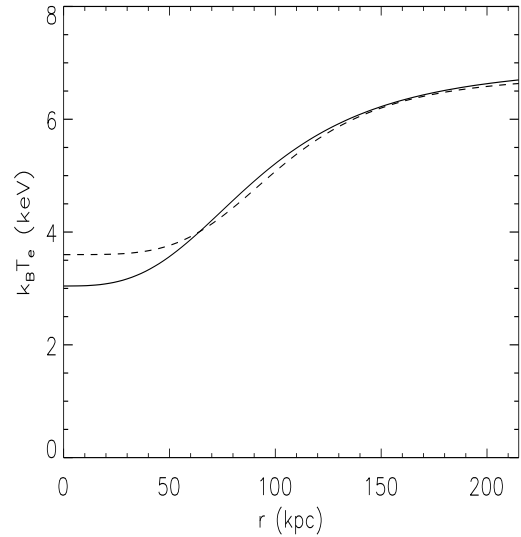,height=8.cm,width=8.cm,angle=0.0}
\end{center}
 \caption{\footnotesize{The deprojected temperature radial profile
 in Perseus derived from the fit to the X-ray data by Chrurazov et
 al. (2003) (solid curve) is compared to the fit to the SZE simulated
 data (dashed curve) shown in Fig. \ref{fig.tproiett_pers}. For the SZE temperature
 profile we use the radial profile given by eq. (\ref{temp.3d}), with
 $k_B T_{ext}=6.8$ keV, $k_B T_{int}=3.6$ keV, $r_c=104$ kpc and $\mu=4.0$.}}
 \label{fig.tdeproiett_pers}
\end{figure}
\begin{figure}[ht]
\begin{center}
 \epsfig{file=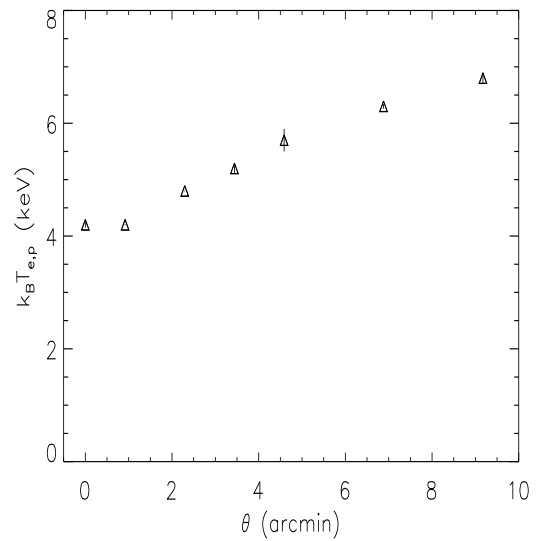,height=8.cm,width=8.cm,angle=0.0}
\end{center}
 \caption{\footnotesize{As Fig. \ref{fig.tproiett_pers} but with
 uncertainties assumed at the level of 0.1\% on the sampled SZE data points
(see Fig. \ref{fig.campionamenti_pers}).}}
 \label{fig.tproiett_pers2}
\end{figure}
The deprojection procedure of the best-fit radial temperature
profile is done by using Eq. (\ref{temp.3d}) setting the values
$k_B T_{ext}=6.8$ keV (the best-fit projected temperature at 200
kpc), and by using as prior constraints the information on
electrons radial profile provided by X-rays observations (see
Sect.4); for an experimental uncertainty of 1\%, such procedure
yields values $k_B T_{int}=3.6\pm1.5$ keV, $r_c=104\pm45$ kpc,
$\mu=4.0\pm2.4$, with $\chi^2=0.1$ for 4 d.o.f.
Fig. \ref{fig.tdeproiett_pers} shows the comparison between the
deprojected temperature profile found with our method and the
best-fit deprojected temperature profile derived by Churazov et
al. (2003). The two curves are quite similar except for the inner
region where the IC gas is cooler and the sensitivity of the SZE
spectrum to low temperatures (through the effect of relativistic
corrections) is lower.
We have also verified that assuming an uncertainty of 0.1\% in the
experimental data, the best-fit values do not change, while the
relative errors are considerably reduced (see Fig.
\ref{fig.tproiett_pers2}).

\subsection{The Ophiuchus cluster}

The Ophiuchus cluster is, at first approximation, an isothermal
cluster with $kT\sim9.9$ keV (Watanabe et al. 2001).
Moreover, being Ophiuchus a quite hot cluster, relativistic
effects in the SZE spectrum are expected to be more prominent and
this allows to obtain a better estimate of both the cluster
temperature and density.

\begin{figure}[ht]
\begin{center}
 \epsfig{file=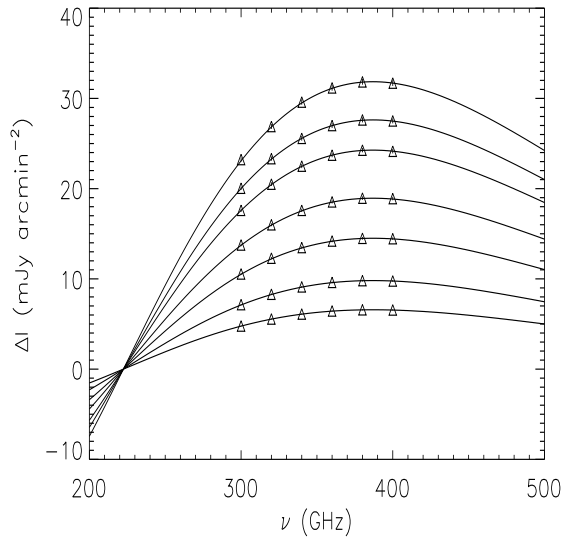,height=8.cm,width=8.cm,angle=0.0}
\end{center}
 \caption{\footnotesize{The thermal SZE spectrum for the Ophiuchus cluster
 is shown at various projected radii of 0, 35, 70, 120, 200, 300 and 500 kpc
 (from top to bottom). The assumed uncertainties on the SZE data
 points is of 1\%. }}
 \label{fig.campionamenti_oph}
\end{figure}

The SZE signal produced in Ophiuchus,
calculated as described in Sect.2, has been sampled
at six frequencies between 300 and 400 GHz at seven
different radii: 0, 35, 70, 120, 200, 300 and 500 kpc (see
Fig.\ref{fig.campionamenti_oph}), assuming experimental
uncertainties of 1\%. The minimum radius at which we sampled the
radial profiles has been chosen taking into account that at the
Ophiuchus distance ($z=0.028$), 1 arcmin corresponds to $\sim$ 34
kpc.

\begin{figure}[ht]
\begin{center}
 \epsfig{file=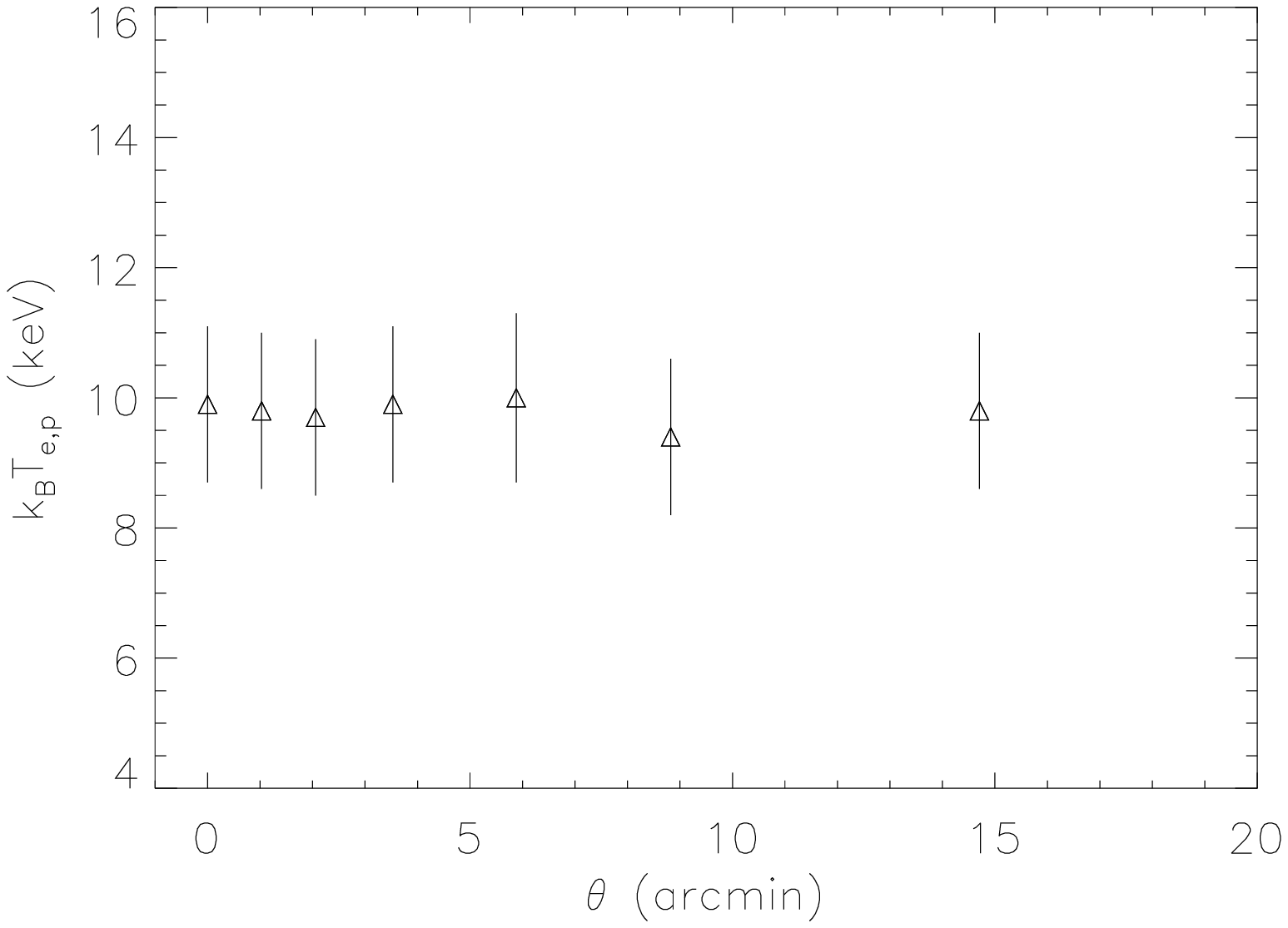,height=8.cm,width=8.cm,angle=0.0}
 \epsfig{file=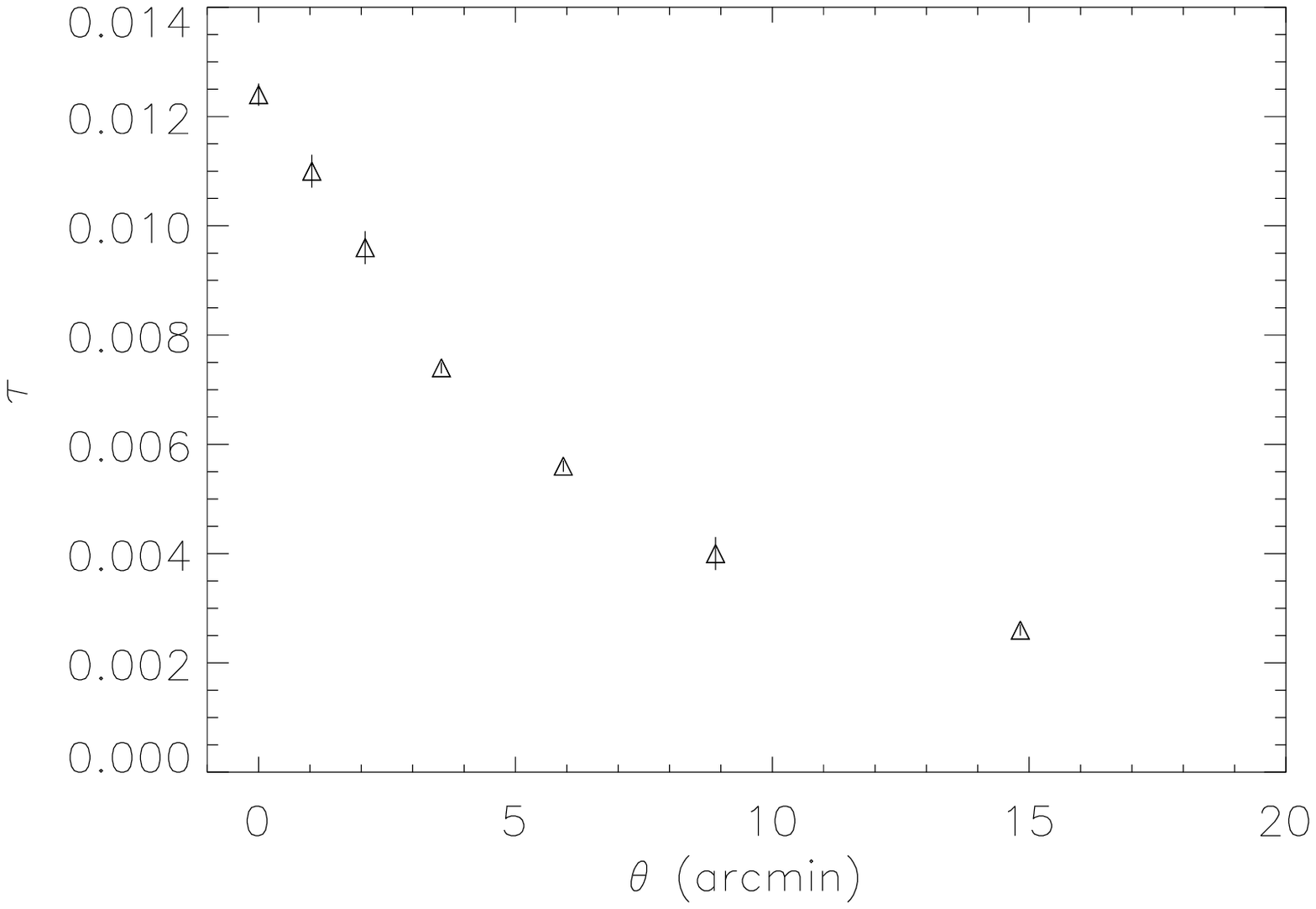,height=8.cm,width=8.cm,angle=0.0}
\end{center}
 \caption{\footnotesize{The radial profile of the projected
 temperature (upper panel) and of the IC gas density (lower panel)
 as derived from the fit to the SZE spectra for Ophiuchus is shown
 as a function of the projected radius (in arcmin).
 We use SZE data points in the frequency range 300--400 GHz with uncertainties of 1\%
 (see Fig. \ref{fig.campionamenti_oph}).}}
 \label{fig.tproiett_oph}
\end{figure}

Fig.\ref{fig.tproiett_oph} shows the radial profiles of the
projected temperature and optical depth obtained from the fit to
SZE data. These results coherently indicate that the cluster temperature is
constant with radius, within the errors, while the radial density
profile can be derived from the optical depth radial profile by
using the following model
\begin{equation}
n_e(r)=n_{e0}\left[ 1+ \left(\frac{r}{r_c}\right)^2 \right]^{-q_{th}}.
\label{eq.densth}
\end{equation}
Therefore, we note that in a cluster with constant temperature
profile, the deprojection procedure can be made by using only
optical depth results without assuming any prior constraint taken
from X-ray data. Hence, the thermal gas parameters derived from
SZE and X-ray data can be directly compared.

\begin{figure}[ht]
\begin{center}
 \epsfig{file=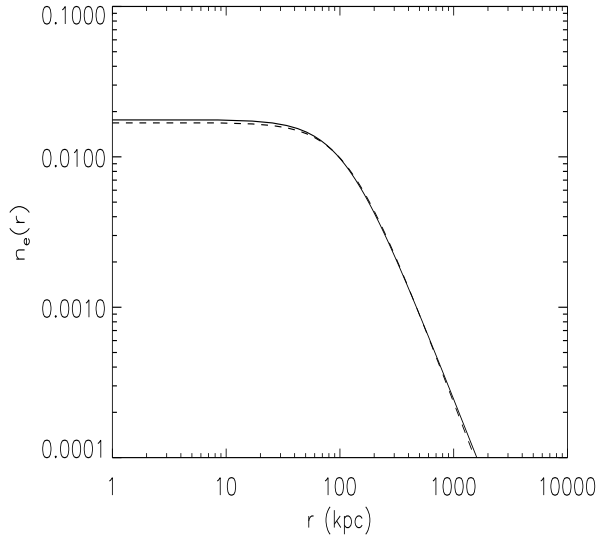,height=8.cm,width=8.cm,angle=0.0}
\end{center}
 \caption{\footnotesize{The IC gas density profile of Ophiuchus as
 derived from X-ray observations (solid curve) is compared with
 the one derived from the optical depth radial profile shown in
Fig.\ref{fig.tproiett_oph} (dashed line) obtained from SZE
simulated observations (see text for details).}}
 \label{fig.dens_oph}
\end{figure}

The values derived from X-ray measurements are
$n_{e0}=1.77\times10^{-2}$ cm$^{-3}$, $r_c=108$ kpc and
$q_{th}=0.96$ (Johnston et al. 1981; Watanabe et al. 2001). The
values we derive from the fit to the simulated SZE signals are
$n_{e0}=(1.7\pm0.2)\times10^{-2}$ cm$^{-3}$, $r_c=118\pm2$ kpc and
$q_{th}=1.00\pm0.07$. The comparison between the best-fit profile
to X-ray data and the one to the SZE data is shown in
Fig.\ref{fig.dens_oph}. We can see that the density profile
derived from SZE measurements is very similar to that obtained
from X-ray observations.

\section{Discussion and conclusions}

We derived in this paper a general formalism for describing the
thermal SZE in galaxy clusters with a non-uniform temperature
profile that can be applied to both cool-core clusters and
non-cool core cluster with an isothermal or non-isothermal
temperature structure.

In order to extract the temperature profile information from
spatially resolved, spectroscopic SZE observations, we derive an
inversion technique through which the electron distribution
function can be extracted using a wide frequency  range coverage
of the SZE signal.
We present extensive results of the fitting procedure used to
extract the cluster temperature from a set of simulated spatially
resolved spectroscopic SZE observations in different bands of the
spectrum, from 100 to 450 GHz, and we discuss the optimal
experimental and theoretical strategy.

The results of our analysis for three different cluster prototypes
(A2199 with a low-temperature cool core, Perseus with a relatively
high-temperature cool core, Ophiuchus with an isothermal
temperature distribution) provide the required precision of the
SZE observations and the optimal frequency bands for  a
determination of the cluster temperature similar or better than
that obtainable from X-ray observations. The precision level of
SZE-derived temperature is also discussed for the  outer regions
of clusters, an issue that is particularly relevant for accurate
mass determination of galaxy clusters.\\
We also study the possibility to extract from our method the
parameters characterizing the non-thermal SZE spectrum of the
relativistic plasma contained in various astrophysical
environments: the lobes of radio galaxies as well as the
relativistic electrons co-spatially distributed  with the thermal
plasma in galaxy clusters with non-thermal phenomena (e.g. radio
halos and/or hard X-ray excesses).

Based on our results, we conclude that the method we present here
is the only method, so far, for SZE observations that is able to
extract the crucial parameters of the cluster atmospheres (i.e.
their temperature, density and additional non-thermal components)
using only a single observational technique, i.e. spatially
resolved spectroscopic SZE observations.

Other methods to extract the cluster temperature profile by using
SZE observations have been presented (Holder \& Loeb 2004), but
they make use of a combination of the thermal SZE (which measures
the electron pressure distribution) and of radio observations of
the highly polarized scattered radiation coming from active
galaxies associated to the cluster (which depends on the cluster
electron density distribution). Holder \& Loeb pointed out that
current instruments should allow us to reach accuracy of the
mass-weighted cluster temperature profiles of order of $\sim 1$
keV, but under the assumption that the central radio source is
steady over several million years. However, variable or beamed
sources (like radio galaxies or jets/lobes of AGNs associated to
the cluster) will leave observable signatures in the scattered
emission.
Therefore, this method could allow, in principle (e.g. with $\mu$K
sensitivity observations to polarized emission), to measure the
age of the central source by finding an edge to the polarized
emission.\\
We noticed in this context, that combining this method with the
independent temperature measurements from SZE spectral studies
(that we discuss here) would allow to measure the time evolution
of the central radio source.

In conclusion, our study shows that the next generation SZE
experiments with spectroscopic capabilities, like those using FTS
spectrometers with imaging capabilities (e.g. SAGACE, {\it
http://oberon.roma1.infn.it/sagacemission/}, and/or MILLIMETRON),
can provide precise temperature distribution over large radial
distances for galaxy clusters even out to substantial redshift.
This will allow to use SZE observations of clusters to better
understand the physics of cluster atmospheres, to better
reconstruct their total mass content and, eventually, to use
clusters of galaxies as reliable astrophysical and cosmological
probes.

\begin{acknowledgements}
We acknowledge stimulating discussions with P.de Bernardis and the
SAGACE team.
%
\end{acknowledgements}

\appendix

\section{Comparing our results to the covariant formalism derivation of the SZE.}

We compare here the SZE calculated according to the formalism of
Wright (1979; W79) with the one calculated according to the
derivation of B\oe hm \& Lavalle (2009; BL09). For this last case,
we refer to their eqs. (18)--(22), derived in the Thomson limit,
that are therefore directly comparable with those derived by W79.

According to BL09, the correct expression of the SZE in the
relativistic covariant formulation, is:
\begin{equation}
\Delta I_\gamma (E_k) =   I_\gamma^{\rm in}(E_k) - I_\gamma^{\rm out}(E_k).
\label{eq:I_in_I_out}
\end{equation}
The averages over the angles of the two contributions are given by
the following expressions
\begin{equation}
\widehat{I}_\gamma^{\rm out}(E_k) = 2\; {\cal K}\; \tau_{\rm nr} \;
I_\gamma^0(E_k),
\end{equation}
where $ {\cal K}\rightarrow 1$ in the Thomson limit, and
\begin{eqnarray}
\widehat{I}_\gamma^{\rm in}  (E_k)&=& 2 \; \tau_{\rm nr} \int dp
\tilde{f}_e(E_p) \nonumber\\ &\times &\int d\mu' \int d\mu \
\frac{{\cal F}(\beta,\mu,\mu')}{\sigma_T} \; I_\gamma^0(t E_k)\;.
\label{eq:I_in3}
\end{eqnarray}
Here, the function $\tilde{f}_e(E_p)$ is normalized as to give
$\int d^3\vec{p}/(2\pi)^3\tilde{f}_e(E_p)=1$, where
$t=(1-\beta\mu)/(1-\beta\mu')$, and the function ${\cal F}$ is
given by:
\begin{eqnarray}
{\cal F}(\beta,\mu,\mu') &\equiv & \frac{\beta^2 m^2}{(2\pi)^3}
\frac{3\sigma_{\rm T}}{16}
\; \frac{(1-\beta\mu')}{ (1-\beta\mu)^2}  \nonumber\\
&\times &  \bigg\{ 2 - 2 K(1-\mu\mu') + K^2 \Big\lbrack (1-\mu\mu')^2\nonumber\\
& & + \frac{1}{2} (1-\mu^2)(1-\mu'^2) \Big\rbrack \bigg\}  \;,
\label{eq:def_fmumup}
\end{eqnarray}
where $K=[\gamma^2(1-\beta\mu)(1-\beta\mu')]^{-1}$.

For the quantitative comparison of the two approaches, we have
used the case of an electron population with a double power-law
spectrum:
\begin{equation}
f_e(p)=A\left\{ \begin{array}{cc}
p^{-s_1} & p_1\leq p \leq p_{break} \\
p_{break}^{-s_1}(p/p_{break})^{-s_2} & p>p_{break}
\end{array} \right.
\label{eq.spettroele}
\end{equation}
where $p=\beta\gamma$,  that is normalized as to give
$\int_0^\infty f_e(p) dp=1$. We use parameters $s_1=0.1$,
$s_2=3.0$, $p_1=1$ e $p_{break}=100$.\\
We notice that, with the use of the a-dimensional momentum $p$ and
the previous normalization, the expression for the function ${\cal
F}(\beta,\mu,\mu')$ in eq. (\ref{eq:def_fmumup}) slightly changes:
for consistency it must be multiplied by the factor
$(2\pi)^3/(\beta^2 m^2 \gamma^2)$, as derived by using eqs.  (16),
(17) and (53) in BL09.

\begin{figure}[ht]
\begin{center}
 \epsfig{file=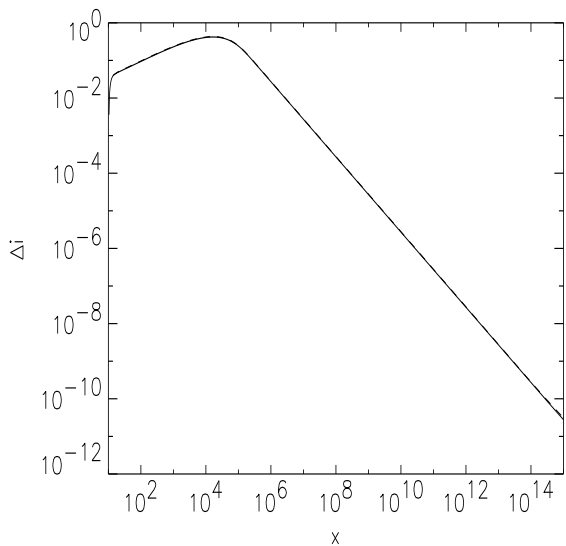,height=8.cm,width=8.cm,angle=0.0}
\end{center}
 \caption{\footnotesize{The SZE spectrum calcolated following the method of W79
 (solid line) and following the method of BL09 (dashed line).
 The horizontal axis shows the a-dimensional frequency $x=(h\nu )/(k_B T_0)$
 while the vertical axis shows the SZE brightness change $\Delta i(x)$
 in units of $2(k_B T_0)^3/(hc)^2$. No difference between the two approaches appears.}}
 \label{fig.boehm.spettro}
\end{figure}

Fig. \ref{fig.boehm.spettro} shows the SZE spectrum plotted over a
wide frequency range, $x=(h\nu )/(k_B T_0)$ from $x=10$ ($\nu \sim
570$ GHz) up to $x=10^{15}$ ($h\nu \sim 2.35 \times10^2$ GeV)
calculated up to first order for $\tau=1$ according to W79
(continuous curve) and that calculated according to BL09 (dashed
line). As is clearly recognized, the two spectra are
indistinguishable over the whole frequency range.

\begin{figure}[ht]
\begin{center}
 \epsfig{file=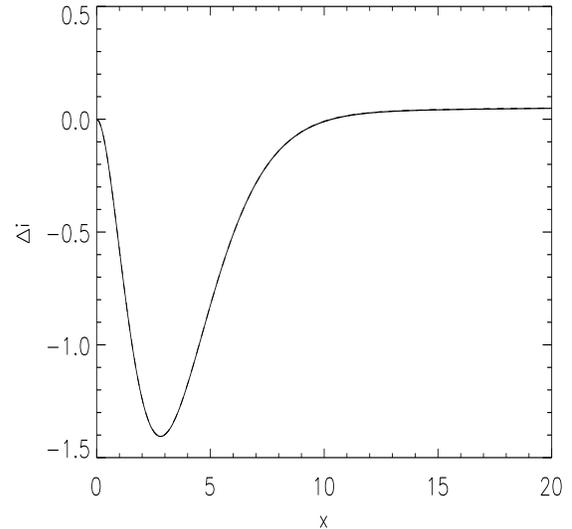,height=8.cm,width=8.cm,angle=0.}
\end{center}
 \caption{\footnotesize{Same as Fig. \ref{fig.boehm.spettro} but
 with a zoom in the microwave frequency range at which the SZE
 telescope operate. Also in this low-frequency range,
 there is no difference between the two approaches.}}
 \label{fig.boehm.micro}
\end{figure}

Fig. \ref{fig.boehm.micro} shows a blow-up of the SZE spectra in
the microwave region  ($x=0  - 20$). Also in this region we must
conclude that the results obtained with the two formalisms are
indistinguishable.

\begin{figure}[ht]
\begin{center}
 \epsfig{file=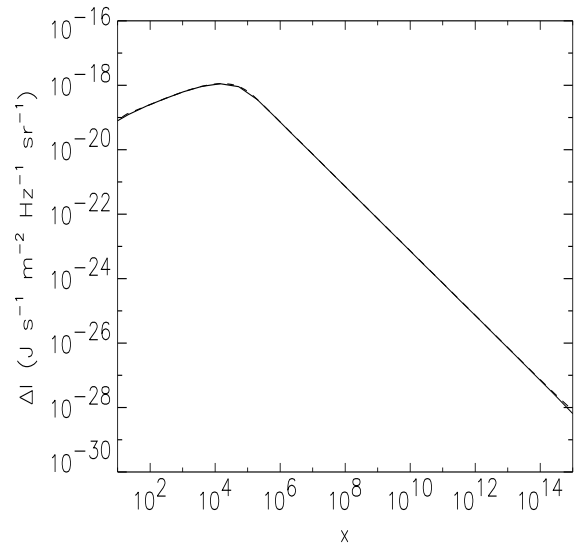,height=8.cm,width=8.cm,angle=0.0}
\end{center}
 \caption{\footnotesize{The comparison of the ICS emission
 evaluated following the standard computational approach with the
 inclusion of the Klein-Nishina cross-section (solid line)
 and the one evaluated following the method of BL09
 in the Thomson approximation (dashed line). No difference appears in the two
 approaches up to a frequency $x \sim 10^{15}$.}}
 \label{fig.boehm.ics}
\end{figure}

Fig. \ref{fig.boehm.ics} shows the comparison between the ICS
emission (against the CMB photons) along the line of sight
calculated in the approach of BL09 and in the standard approach
(see e.g. Colafrancesco et al. 2005, see also Longair 1993) using
also the Klein-Nishina cross-section, and assuming an electron
spectrum as in eq.(\ref{eq.spettroele}), with a density and
spatial distribution such that $\tau=1$.
The two different calculations provide quite similar results,
apart for the frequency range $x > 10^{15}$, at which the
Klein-Nishina cross-section has a non-negligible effect (it
decreases, in fact, the Compton scattering efficiency at high
energies, see e.g. Fargion \& Salis 1998).

Based on the previous results, we can conclude that:\\
i) for the SZE spectrum in the microwave--mm region the formalism
of W79 can be used also in the case of relativistic electrons
since it provides the same results of the covariant formalisms of
BL09. In addition, the W79 formalism, that uses a more simple
expression requires a much shorter computing time w.r.t. the BL09
formalism.\\
ii) for the calculation of the ICS emission at high frequencies,
the standard formalism is completely adequate since also in this
case there is no difference w.r.t. the covariant formalism
results. In addition, the standard formalism requires much shorter
computing time and allows also the use of the Klein-Nishina
cross-section effect without introducing additional complications
in the numerical computation, as occurs - on the contrary - in
eq.(46) of BL09.

Recently Nozawa \& Kohyama (2009) analyzed the covariant formalism
of the Sunyaev-Zeldovich effect for the thermal and nonthermal
distributions and derived the frequency redistribution function
identical to the W79 method assuming the smallness of the photon
energy (in the Thomson limit). They also derive the redistribution
function in the covariant formalism in the Thomson limit. These
authors have shown that two redistribution functions are
mathematically equivalent in the Thomson limit which is fully
valid for the cosmic microwave background photon energies.

To summarize, the W79 and the covariant formalisms to calculate
the SZE are fully equivalent, contrary to previous erroneous
claims (see BL09), and the large advantage given by the standard
calculations based on the W79 approach are due to the much shorter
computing times.

\end{document}